\documentclass[prl,twocolumn,floatfix,showpacs ]{revtex4-1}
\usepackage{amsmath}
\usepackage{amssymb}
\usepackage{graphicx}
\usepackage{dcolumn}
\usepackage{bm}
\usepackage[colorlinks=true,linkcolor=blue,anchorcolor=blue, citecolor=cyan,urlcolor=cyan]{hyperref}
\usepackage[mathlines]{lineno}
\usepackage{ulem}
\usepackage{epstopdf}

\newcommand{\ks}[1]{|#1 \rangle}
\newcommand{\bs}[1]{\langle #1 |}

\begin{document}
\title{Twisted Magnetic Van der Waals Bilayers: An Ideal Platform for Altermagnetism}
\author{Yichen Liu}
\thanks{These authors contributed equally to this work.}
\author{Junxi Yu}
\thanks{These authors contributed equally to this work.}
\author{Cheng-Cheng Liu}
\email{ccliu@bit.edu.cn}
\affiliation{Centre for Quantum Physics, Key Laboratory of Advanced Optoelectronic Quantum Architecture and Measurement (MOE), School of Physics, Beijing Institute of Technology, Beijing 100081, China}
\begin{abstract}
We introduce a universal methodology for generating and manipulating altermagnetism in two-dimensional (2D) magnetic van der Waals (MvdW) materials through twisting. We find that a key in-plane 2-fold rotational operation can be achieved in a twisted bilayer of any 2D MvdW material, which takes one of all five 2D Bravais lattices, thereby inducing altermagnetism. By choosing the constituent MvdW monolayer with specific symmetry, our approach can tailor altermagnetism of any type, such as $d$-wave, $g$-wave, and $i$-wave. Furthermore, the properties of our twisted altermagnetic materials can be easily engineered. Taking a transition-metal oxyhalide VOBr as an example, we find that by tuning the twist angle and Fermi level a giant spin Hall angle can be obtained, much larger than the experimentally reported. This approach establishes a general, robust, and adjustable platform to explore altermagnetism, and provides a new efficient way to generate and manipulate the spin current.
\end{abstract}
\maketitle
\textit{Introduction.---}Recently, a type of collinear crystal-symmetry compensated magnetic order called altermagnetism~\cite{smejkalConventionalFerromagnetismAntiferromagnetism2022,smejkalEmergingResearchLandscape2022}, which uniquely combines the zero net magnetization and nonrelativistic spin splitting in the Brillouin zone (BZ), has emerged as an exciting research landscape~\cite{fengAnomalousHallEffect2022,baiEfficientSpintoChargeConversion2023,zhouCrystalThermalTransport2024a,ouassouDcJosephsonEffect2023,liMajoranaCornerModes2023,zhuTopologicalSuperconductivityTwodimensional2023,zhangFinitemomentumCooperPairing2024,smejkalCrystalTimereversalSymmetry2020a,leeBrokenKramersDegeneracy2024a,mazinAltermagnetismMnTeOrigin2023a,kriegnerMagneticAnisotropyAntiferromagnetic2017,loveseyTemplatesMagneticSymmetry2023,yuanGiantMomentumdependentSpin2020a,liuInverseMagnetocaloricEffect,brekkeTwodimensionalAltermagnetsMinimal2023,sodequistTwodimensionalAltermagnetsHigh2024,sodequistTwodimensionalAltermagnetsHigh2024,mazinInducedMonolayerAltermagnetism2023,guoPiezoelectricAltermagnetismSpinvalley2023}. Such a combination results in many fascinating properties~\cite{fengAnomalousHallEffect2022,baiEfficientSpintoChargeConversion2023,zhouCrystalThermalTransport2024a,ouassouDcJosephsonEffect2023,liMajoranaCornerModes2023,zhuTopologicalSuperconductivityTwodimensional2023,zhangFinitemomentumCooperPairing2024} including crystal Hall effect, spin current, giant crystal thermal transport, etc., in some altermagnetic materials, such as RuO$_2$, MnTe, and MnF$_2$.
However, research on altermagnetism has been mainly focused on three-dimensional systems, with less attention paid to two-dimensional (2D) systems. Therefore, a universal approach to generally constructing 2D altermagnetism has recently become an urgent need in the community.

Since the discovery of ferromagnetic order in monolayer CrI$_3$ in 2017~\cite{huangLayerdependentFerromagnetismVan2017}, there has been a surge in interest in magnetic van der Waals (MvdW) materials, leading to the identification of abundant 2D experimentally-prepared magnetic materials, including ferromagnetic CrX$_3$ (X = Cl, Br)~\cite{bedoya-pintoIntrinsic2DXYFerromagnetism2021,zhangDirectPhotoluminescenceProbing2019}, CrBrS~\cite{leeMagneticOrderSymmetry2021}, MnSe~\cite{aaproSynthesisPropertiesMonolayer2021}, and MnBi$_2$Te$_4$~\cite{MnBiTe2019_Nature,zhangTopologicalAxionStates2019prl,li2019intrinsicSA}, all of which have interlayer antiferromagnetic order, and antiferromagnetic materials, such as transition-metal thiophosphates MPS$_3$ and selenophosphates MPSe$_3$~\cite{oharaRoomTemperatureIntrinsic2018} (M=Fe,Mn,Cr), along with FeTe~\cite{chandraElasticPropertiesMono2010,gomezEffectsSubstitutingSe2010,ciechanMagneticPhaseTransitions2013} and CrTe$_3$~\cite{mcguireAntiferromagnetismVanWaals2017}. The properties of vdW materials can be adjusted through various means~\cite{naumisElectronicOpticalProperties2017,caoUnconventionalSuperconductivityMagicangle2018,andreiGrapheneBilayersTwist2020,xuMoireFlatBands2021,viznersternInterfacialFerroelectricityVan2021,wuLargeSlidingRegulation2023,wuSlidingFerroelectricity2D2021}, such as pressure, slide, electric field, and notably twist, which has received widespread
attention as a new degree of freedom~\cite{caoUnconventionalSuperconductivityMagicangle2018,andreiGrapheneBilayersTwist2020}. Taking advantage of the rich 2D MvdW materials, it will become feasible to construct 2D altermagnets and engineer their physical properties.

\begin{figure*}[t]
    \centering
    \includegraphics[width=0.98\textwidth]{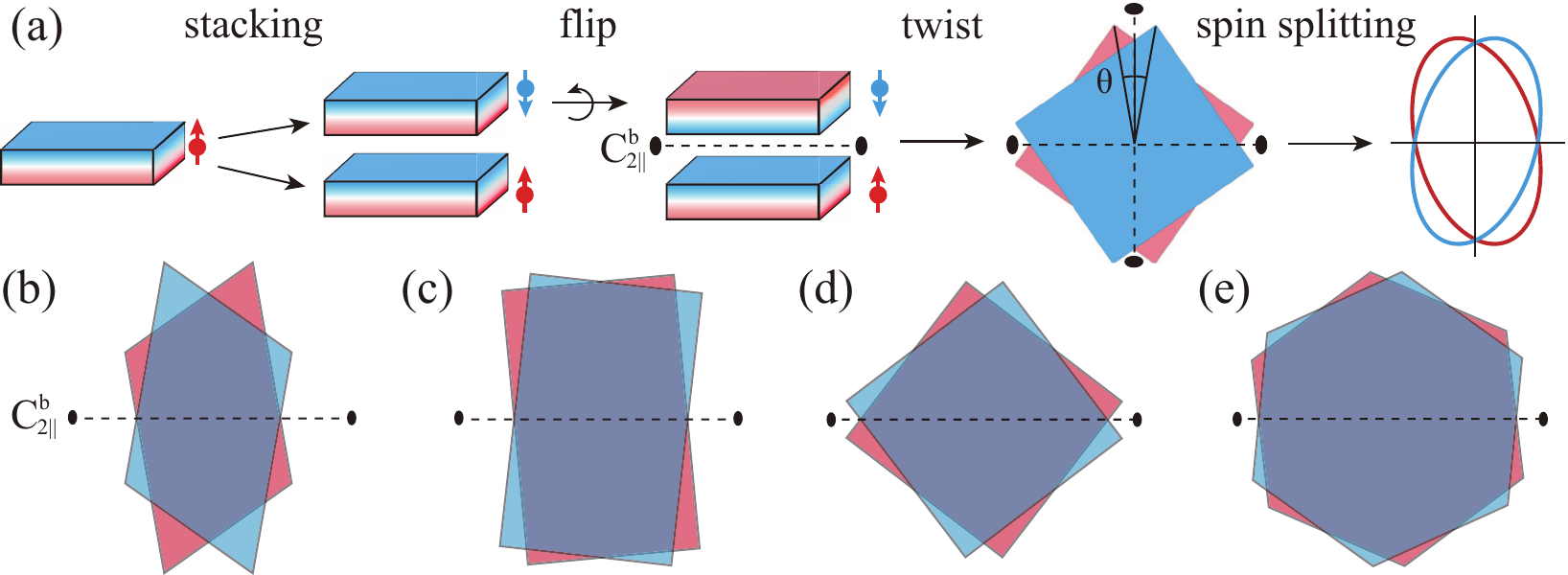}
    \caption{\label{fig:illu}(a) Illustration of a general route to altermagnetism in twisted magnetic vdW materials. Initially, monolayer MvdW materials are stacked into a bilayer structure that exhibits antiferromagnetic order. Subsequently, the upper layer is flipped to introduce an in-plane 2-fold axis $C_{2\parallel}^b$. Then the upper and lower layers are rotated by $\frac{\theta}{2}$ and $-\frac{\theta}{2}$ respectively, which breaks inversion symmetry and gives rise to spin splitting, as depicted in the final stage. (b-e) Through the process shown in (a), the key $C_{2\parallel}^b$ operation marked by dashed lines can be achieved in a twisted bilayer of any 2D MvdW material, which takes one of all five 2D Bravais lattices, thereby inducing altermagnetism.  A twisted bilayer of the (b) oblique lattice with $C_2$ point group, (c) rectangular/centered-rectangular lattice with $D_2$ point group, (d) square lattice with $D_4$ point group, and (e) hexagonal lattice with $D_3$/$D_6$ point group.}
\end{figure*}

In this Letter, we propose a universal approach to realize altermagnetism in MvdW materials via twist and demonstrate its universality, stability, and tunability. By symmetry analysis and the construction of twisted tight-binding (TB) models, we show that our approach can generate $d$-wave, $g$-wave, and $i$-wave altermagnetism, and is further corroborated by the density-functional (DFT) calculations. Furthermore, the physical properties of altermagnets generated from our method are adjustable by twist angle, strain, electric field, etc. As an example, we investigate $d$-wave altermagnetism in VOBr and find that by varying the twist angle and gate voltage, the spin Hall angle (SHA) can take 1.4, significantly surpassing the previous experimental reports. Such giant SHA in the twisted magnetic vdW materials allows for highly efficient spin-current generation. Given its universality and tunability, our approach builds a flexible and ideal platform for altermagnetism.

\textit{Approach \& symmetry analysis.---}
To generate altermagnetism, two conditions must be met~\cite{smejkalConventionalFerromagnetismAntiferromagnetism2022}: Firstly, the system must exhibit the compensated collinear magnetic order. Secondly, the sublattices with opposite spins should be connected via a rotation symmetry. The rotation in 2D materials can be divided into out-of-plane and in-plane. Out-of-plane rotation varies depending on the specific material and thus lacks commonality. In contrast, regardless of the symmetry of the material, an in-plane two-fold rotation operator can always be achieved by stacking the bilayer structure. Therefore, using the in-plane rotation to construct altermagnetism is a general approach with the only condition that the MvdW material has interlayer antiferromagnetic order, which is commonly found in the MvdW bilayers~\cite{burchMagnetismTwodimensionalVan2018c,huangLayerdependentFerromagnetismVan2017,feiTwodimensionalItinerantFerromagnetism2018,liusupplemental}.

As shown in Fig.~\ref{fig:illu}(a), the approach of generating altermagnetism starts with stacking into a bilayer structure with interlayer antiferromagnetic ordering. To fulfill the second condition for generating altermagnetism, the upper layer is flipped, introducing a key in-plane rotation operation $C_{2\parallel}^b$. A twist operation follows with inversion symmetry (if any) broken. It is worth noting that after the twist, the $C_{2\parallel}^b$ remains. The $C_{2\parallel}^b$ operation, which connects two layers with opposite spins and plays a key role in generating altermagnetism, is always preserved in our twisted MvdW bilayers. See details in Supplemental Material (SM)~\cite{liusupplemental}. From the perspective of the spin group, this configuration consistently exhibits $[C_{2}][C_{2\parallel}^b]$ symmetry, where the former $C_{2}$ reverses the spin orientation in spin space and $C_{2\parallel}^b$ flips the layers in real space. Our approach applies to all interlayer antiferromagnetic vdW materials, whose monolayer can take any 2D Bravais lattice, demonstrating the broad applicability and universality of the route to altermagnetism.

In our approach, the in-plane rotational operation $C_{2\parallel}^b$, which can always exist in the above-mentioned twisted MvdW systems, plays a decisive role in the generation of altermagnetism, while the out-of-plane rotational symmetry will determine the symmetry of spin splitting, i.e., $d$-wave, $g$-wave or $i$-wave. The $C_{2\parallel}^b$ rotation links sublattices with antiparallel spins, causing spin degeneracy on the invariant paths of $C_{2\parallel}^b$ within BZ. As depicted in Figs.~\ref{fig:illu}(b)-(e), the twisted systems of different Bravais lattices possess distinct $C_{2\parallel}^b$ axes with diverse spin-degenerate paths and thus can induce various forms of spin splitting, including $d$-wave, $g$-wave, and $i$-wave altermagnetism.

\begin{figure}[t]
    \centering
    \includegraphics[width=0.48\textwidth]{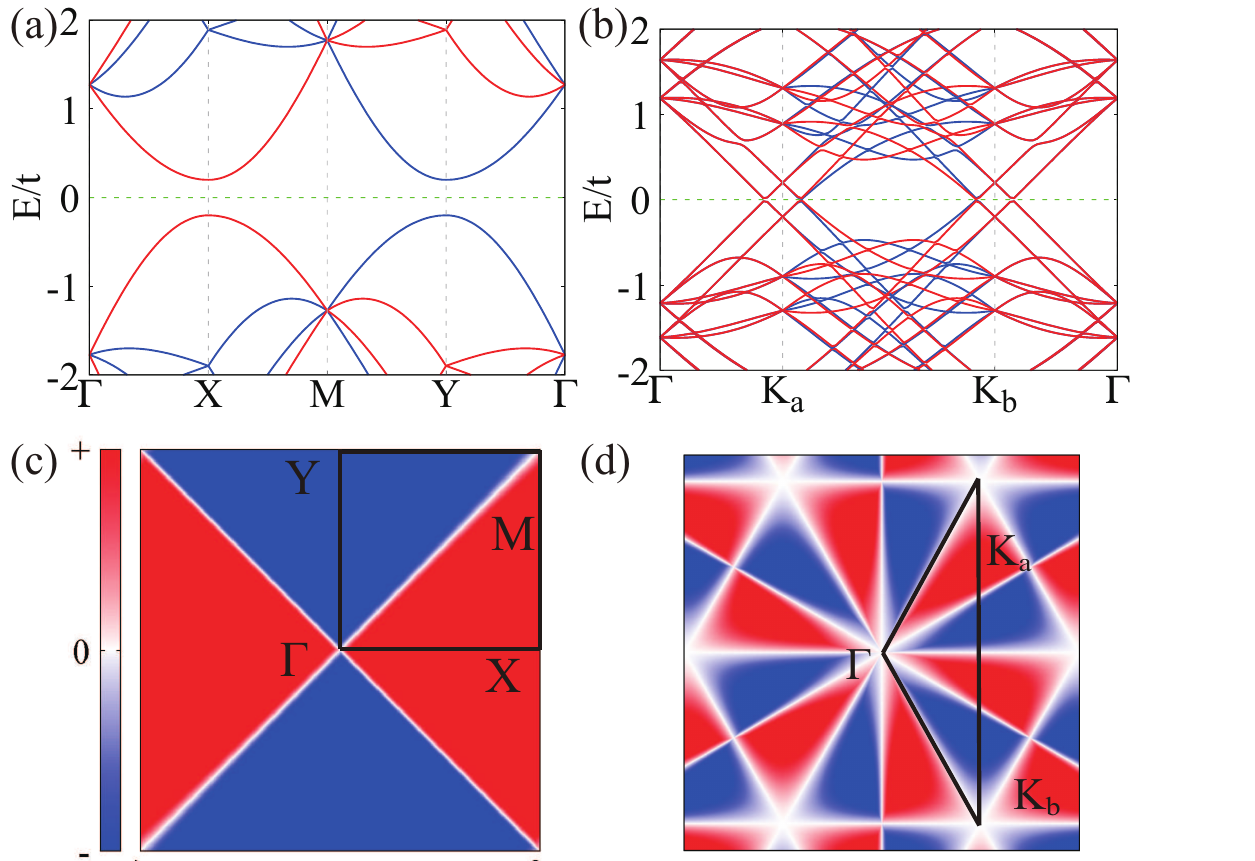}
    \caption{\label{fig:model}(a) The band structure of the twisted model of $h_1$ with $t_1=-0.5 t_2=t$, $M=3.2t$, and $\bm a_1 \perp \bm a_2 $. (b) The band structure of the twisted model of $h_2$, wherein $\delta=0$ and $M=0.2 t$. (c) (d) Their respective spin splitting with $d$- and $i$-wave symmetry. The red and blue areas denote positive and negative spin splitting of the valence band, respectively, while the white indicates regions of spin-degeneracy.}
\end{figure}

\textit{Model.---}We now demonstrate the application of our method using the twisted TB model, and explore how the symmetry of MvdW materials affects altermagnetism. We develop monolayer models for both oblique and hexagonal lattices, capable of exhibiting $E$, $C_{2z}$ and $C_{4z}$ symmetries, while $C_{3z}$ and $C_{6z}$ symmetries, respectively, by adjustable parameters
\begin{eqnarray}
h_1(\bm k)&&=t_1\cos(\bm k\cdot\bm a_1)+t_2\cos(\bm k\cdot\bm a_2),\\
h_2(\bm k)&&=\sum_{i=1}^3 t\left[\cos(\bm k\cdot \bm d_i)\tau_x+\sin(\bm k\cdot\bm d_i)\tau_y\right]+\delta\tau_z.\label{tb:model}
\end{eqnarray}
Here, $h_1$ uniformly describes the Hamiltonian of oblique, rectangular, and square monolayer systems. $\bm a_i$ represent lattice vectors, $t_i$ are hopping. $h_2$ is the Hamiltonian of the hexagonal system with $\bm d_i$ the nearest neighbor vectors, $\tau$ the Pauli matrices labeling the sublattices, $t$ the hopping and $\delta$ the staggered potential. When $t_1$ equals $t_2$ and $\bm a_1\perp \bm a_2$ with equal lengths, $h_1$ has $C_{4z}$ symmetry; whereas when $t_1$ and $t_2$ are not equal, it shows $C_{2z}$ symmetry. As for $h_2$, it displays $C_{6z}$ symmetry when $\delta$ is zero, and in other cases it has $C_{3z}$ symmetry.

As illustrated in Fig.~\ref{fig:illu}(a), we flip the upper layer and rotate the upper and lower layers by $-\frac{\theta}{2}$ and $\frac{\theta}{2}$, respectively, forming a moir\'{e} supercell. The twisted TB model is constructed
\begin{equation}
H=\left(\begin{array}{cc}
C_{2\parallel}^m h(R(-\frac{\theta}{2})\bm k)C_{2\parallel}^m+M s_z & T\\
T^\dag & h(R(\frac{\theta}{2})\bm k)-Ms_z \\
\end{array}\right).
\end{equation}
Where $C_{2\parallel}^m$ is the flip operation, $R$ is the rotation matrix, $T$ represents the interlayer coupling, $s_z$ labels the spin on each layer, and $M$ represents the magnetic moment. The interlayer coupling is added using the Slater-Koster
method~\cite{bistritzerMoireBandsTwisted2011,jungInitioTheoryMoir2014}.

The spin-splitting in the band structures of our twisted TB model is shown in Figs.~\ref{fig:model}(a) and (b). Notably, altermagnetism possesses $[C_2][\mathcal{T}]$ symmetry, where $\mathcal{T}$ represents time-reversal symmetry and $C_2$ inverses the spins in spin space, leading to $H(\bm k)=H(-\bm k)$, which classifies altermagnetism into even-parity forms~\cite{smejkalConventionalFerromagnetismAntiferromagnetism2022}. Corresponding to  $C_2$ and $D_2$ point groups, we obtain $d$-wave altermagnetism (Fig.\ref{fig:model}(c)), $D_4$ group gives rise to $g$-wave~\cite{liusupplemental}, and both $D_3$ and $D_6$ groups lead to $i$-wave altermagnetism~\cite{liusupplemental}, as plotted in Fig.~\ref{fig:model}(d). Thus, by selecting MvdW materials with diverse symmetries, our method can tailor any even-parity waveform of altermagnetism, as shown in Table~\ref{tab:table1}.

\begin{figure}[t]
    \centering
    \includegraphics[width=0.48\textwidth]{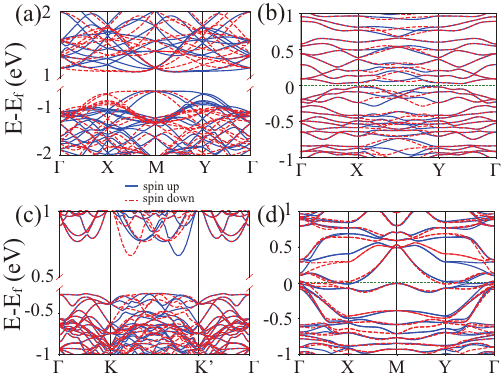}
    \caption{\label{fig:DFT}(a) The band structure of twisted bilayer VOBr with orthogonal lattice and $d$-wave altermagnetism. The twist angle is $48.16^\circ$. Red dashed lines and blue lines represent spin up and down bands, respectively. (b) The band structure of twisted bilayer Co$_2$S$_2$ with tetragonal lattice and $g$-wave altermagnetism. The twist angle is $53.12^\circ$. (c) The band structure of twisted bilayer MnBi$_2$Te$_4$ with $i$-wave altermagnetism. The twist angle is $53.13^\circ$. (d) The band structure of twisted bilayer antiferromagnetic vdW materials FeTe with $d$-wave altermagnetism. The twist angle is $21.79^\circ$.
    The k-path for (a), (b), and (d) is depicted in Fig.~\ref{fig:model}(c), and the k-path for (c) is illustrated in Fig.~\ref{fig:model}(d).
    }
\end{figure}

\textit{Material realization.---}Guided by our generic design principles and combined with DFT calculations, we predict ample altermagnetic candidates, which are classified into rectangular, tetragonal, and triangle systems, respectively~\cite{liusupplemental}. For the rectangular system, the interlayer antiferromagnetic transition-metal oxyhalides MOY (M=Ti, V, Cr; Y=Cl, Br) are considered, which belong to the Pmmn space group and have been synthesized~\cite{miao2DIntrinsicFerromagnets2018b,coicChromiumIIIOxyhalides1981,fengAntiferromagneticFerroelasticMultiferroics2021,wiedenmannMagneticOrderingQuasitwodimensional1983}. As a specific example, we choose VOBr, whose the experimental lattice constants $a=3.775$ \AA\ and $b=3.38$ \AA~\cite{miao2DIntrinsicFerromagnets2018b} are used, to demonstrate the band splitting after twist. The band structure with $d$-wave altermagnetism is shown in Fig.~\ref{fig:DFT}(a).
For the tetragonal system, the PbO-type interlayer antiferromagnetic vdW materials M$_2$X$_2$(M=Mn, Fe, Co, Ni; X=S, Se, Te) ~\cite{shenHighthroughputComputationStructure2022,zhangTwodimensionalMagneticMaterials2021a,torelliHighthroughputComputationalScreening2020} are considered. Here we use Co$_2$S$_2$~\cite{zhangTwoDimensionalCo2S2Monolayer2017}, which is predicted as intralayer ferromagnetism, to show the spin-splitting of $g$-wave altermagnetism. As for the hexagonal lattice, we choose the well-known material MnBi$ _2$Te$_4$ as an example. The twisted system has $i$-wave altermagnetism, as plotted in Fig.~\ref{fig:DFT}(c).

\begin{table}[t]
\caption{\label{tab:table1}%
An ideal platform for altermagnetism by twisting magnetic van der Waals bilayers. Our approach can produce altermagnetism of any type by choosing the MvdW monolayer with specific symmetry. The symmetry of the constituent monolayer, the corresponding altermagnetic types, and the candidates are presented in Columns 1-3, respectively.
}
\begin{ruledtabular}
\begin{tabular}{ccc}
monolayer &  altermagnetism & material \\
symmetry &    symmetry & canditates\\
\colrule
$E$, $C_{2z}$ & $d$-wave & FeTe,MOY\\
& & (M=Ti,V,Cr;Y=Cl,Br)\\
$C_{4z}$ & $g$-wave &  M$_2$X$_2$\\
& & (M=Mn,Fe,Co,Ni;X=S,Se,Te)\\
$C_{3z}$, $C_{6z}$ & $i$-wave & CrI$_3$, MnBi$_2$Te$_4$, NaCl$_2$, etc.
\end{tabular}
\end{ruledtabular}
\end{table}

Our approach only needs the interlayer antiferromagnetic order and has no requirements for the intralayer magnetic order. For instance, FeTe~\cite{chandraElasticPropertiesMono2010,gomezEffectsSubstitutingSe2010,ciechanMagneticPhaseTransitions2013} demonstrates antiferromagnetic order both intralayer and interlayer, and its monolayer maintains $\mathcal{PT}$ symmetry. After the twist, the inversion is destroyed and the system becomes a $d$-wave altermagnet, as shown in Fig.~\ref{fig:DFT}(d). Considering the $\mathcal{PT}$ symmetry inherent in each layer, the magnitude of spin splitting is directly related to the strength of interlayer coupling. In Table~\ref{tab:table1}, we tabulate the various forms of altermagnetism that can be generated through our approach by choosing monolayer MvdW materials with different symmetries and the abundant candidate materials for each form of altermagnetism.

\textit{Tunability with a giant SHA.---}
One key advantage of the altermagnetism we have developed is its tunability of physical properties. Specifically, we can modulate these properties by adjusting the twist angle and applying a gate voltage. To demonstrate this, we focus on the spin current.
The spin-current conductivity can be expressed as~\cite{freimuthSpinorbitTorquesCo2014c}
\begin{equation}
\chi_{\alpha\beta}^\gamma =-\frac{e\hbar}{V\pi}\text{Re}\sum_{\bm k,mn}\frac{\eta^2\bs{u_{n\bm k} }J_{\alpha}^\gamma\ks{u_{m\bm k}}\bs{u_{m\bm k}}v_y\ks{u_{n\bm k}}}{(\mathcal{E}_n^2+\eta^2)(\mathcal{E}_m^2+\eta^2)}.\label{eq:SHE}
\end{equation}
where $\ks{u_{n\bm k}}$ is the Bloch function of the $n$th band, $\bm k$ is the wave vector, $J_{\alpha}^\gamma=\frac{1}{2}\left\{s_\gamma,v_{\alpha}\right\}$ is the spin-current operator, $\mathcal{E}_n=E_f-\varepsilon_{n\bm k}$, $\varepsilon_{n\bm k}$ is the band energy, $E_f$ is the Fermi energy, and $\eta$ is the constant scattering-rate. Due to crystal symmetry constraint~\cite{seemannSymmetryimposedShapeLinear2015}, the conductivity tensor is diagonalized when the system possesses $C_{3z}$, $C_{4z}$, and $C_{6z}$ symmetries, hence the oblique and rectangular systems need to be considered. To eliminate the influence of the material-dependent variable $\eta$, we characterize our results using SHA, denoted as $\Theta_\text{SH}=\frac{2e}{\hbar}\chi_{xy}^z/\chi_{xx}$~\cite{mosendzQuantifyingSpinHall2010,isasaTemperatureDependenceSpin2015a}.

\begin{figure}[t]
    \centering
    \includegraphics[width=0.48\textwidth]{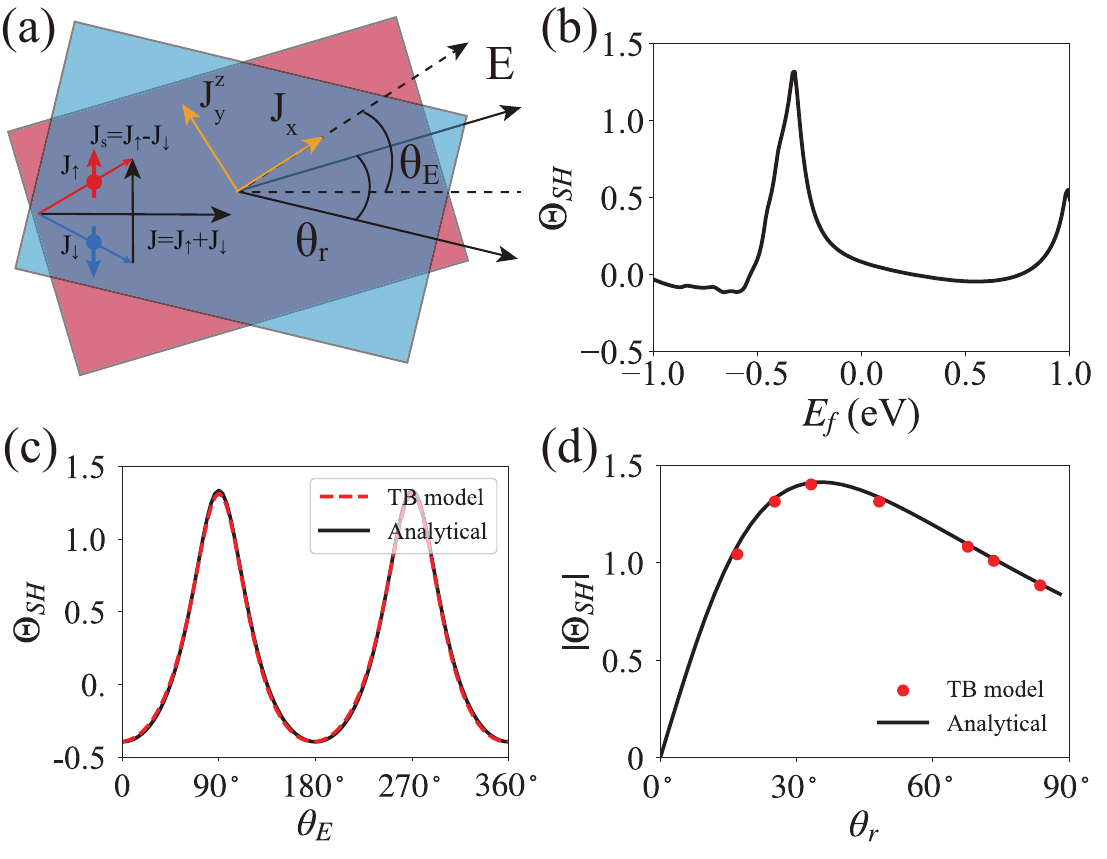}
    \caption{\label{fig:current}(a) A twisted bilayer VOBr simple. The spin current $J_y^z$ and the current $J_x$ are perpendicular and parallel to the direction of the electric field, respectively. The spin current and the current can be obtained by superposing the spin-polarized current generated in a monolayer. The twist angle $\theta_r$ and the azimuth angle of the electric field $\theta_E$ are illustrated. (b) The spin Hall angle changes with Fermi level $E_f$ and $\theta_r=48.16^\circ$ and $\theta_E=90^\circ$. (c) The spin Hall angle changes with $\theta_E$ under $E_f=-0.33$ eV and $\theta_r=48.16^\circ$. (d) The Spin hall angle changes with $\theta_r$ under $E_f=-0.33$ eV and $\theta_E=90^\circ$. The red dots are the results from the twisted TB model with $\theta_r$ at commensurable angles, and the black line is the analytical result from Eq.~(\ref{eq:hall}) with $\chi_{yy}^0/\chi_{xx}^0=0.101.$
    }
\end{figure}

Altermagnetic materials have a spin-current effect due to their spin-polarized Fermi surfaces~\cite{karubeObservationSpinSplitterTorque2022,baiEfficientSpintoChargeConversion2023}. In our twisted systems, due to the spin-layer locking, we can give a more intuitive explanation of the mechanism of spin current. As shown in Fig.~\ref{fig:current}(a), when an electric field is applied, the upper and lower layers will generate opposite spin-polarized currents in different directions.
The currents parallel and perpendicular to the direction of the electric field are denoted as $J_x$ and $J_y^z$, respectively. From the above picture, we can give the conductivity formula of twisted altermagnetism as follows (see details in SM~\cite{liusupplemental})
\begin{subequations}
\label{eq:hall}
\begin{eqnarray}
\chi_{xy}^z&&=2\chi_{xy}^0-\frac{1}{2}\left(\chi_{xx}^0-\chi_{yy}^0\right)\sin\theta_r\cos 2\theta_E,\\
\chi_{xx}&&=\chi_{xx}^0+\chi_{yy}^0+(\chi_{xx}^0-\chi_{yy}^0)\cos\theta_r\cos 2\theta_E.
\end{eqnarray}
\end{subequations}
The $\chi_{xy}^z$ and $\chi_{xx}$ represent the spin Hall conductivity and normal conductivity after twist respectively, and $\chi_{ab}^0$ represents the conductivity of the single layer in the untwisted system. The $\theta_r$ and $\theta_E$ are the twist angle and the angle between the electric field and the $C_{2\parallel}^b$ axis, as illustrated in Fig.~\ref{fig:current}(a). When the monolayer has $C_{2x}$ symmetry,  $\chi_{xy}^0=0$. We obtain the analytic expression for the maximum SHA  with the corresponding optimal twist angle (see details in SM~\cite{liusupplemental})
\begin{equation}
\theta_r^*=\arccos\left(\frac{\chi_{xx}^0-\chi_{yy}^0}{\chi_{xx}^0+\chi_{yy}^0}\right),\ \Theta_\text{SH}^\text{max}=\frac{\chi_{xx}^0-\chi_{yy}^0}{2\sqrt{\chi_{xx}^0\chi_{yy}^0}}.\label{eq:max}
\end{equation}

Here, we use the twisted VOBr with $d$-wave altermagnetism as an example. We first study the change of $\Theta_\text{SH}$ with the Fermi level at a commensurable angle $\theta_r=48.16^\circ$ and $\theta_E=0^\circ$. As shown in Fig.~\ref{fig:current}(b), the SHA reaches the maximum value $\Theta_\text{SH}=1.3$ when $E_f=-0.33$eV, far beyond the best results currently reported in experiments~\cite{isasaTemperatureDependenceSpin2015a,mosendzQuantifyingSpinHall2010,obstbaumTuningSpinHall2016,singhLargeSpinHall2020,wangDeterminationIntrinsicSpin2014}. Next we calculate the conductivity of bilayer VBrO and get $\chi_{yy}^0/\chi_{xx}^0=0.101$, indicating obvious anisotropy. We bring it into Eq.~(\ref{eq:hall}), and compare it with the twisted TB model at $\theta_r=48.16$, $E_f=-0.33$ eV, shown in Fig.~\ref{fig:current}(c). It can be shown that only small differences between the two results, thereby affirming the reliability of Eq.~\eqref{eq:hall}. Subsequently, we explore the variation of the SHA with respect to $\theta_r$. Given the limitation of commensurable angles within a rectangular structure, the twisted TB model with several commensurate angles is considered (see details in SM~\cite{liusupplemental}). The results from the TB model agree with the analytical results derived from Eq.~\eqref{eq:hall}, as shown in Fig.~\ref{fig:current}(d). According to Eq.~\eqref{eq:max}, we find that the SHA reaches its peak at $\theta_r^*=35.27^\circ$ with a maximum value of $\Theta_\text{SH}^\text{max}=1.4$.

\textit{Conclution.---} In this work, we provide a general route to generating and engineering altermagnetism in the bilayer of MvdW materials through the twist. Our approach offers a stable and controllable ideal platform for altermagnetism, provides rich possibilities for research in this field, and has three key advantages: (1) Universality: Both the flip operation and the twist operation are independent of constituent monolayer symmetry, so our method can be applied to any MvdW material with interlayer antiferromagnetic order, such as CrI$_3$, MnBi$_2$Te$_4$, FeTe, etc., allowing a wide selection of materials and the generation of various forms of altermagnetism, such as $d$/$g$/$i$-wave. (2) Robustness: Interlayer slide $\bm\tau$ does not affect the existence or the even-parity form of altermagnetism, which greatly improves the possibility of experimental implementation of the scheme~\cite{liusupplemental}. (3) Tunability: Benefiting from the tunability of 2D materials, our approach allows direct engineering of the physical properties of the twisted bilayer through twist angle, doping, strain, pressure, etc. Taking a rectangular transition-metal oxyhalide VOBr as an example, we have discovered that modifying the twist angle and doping can result in a giant SHA significantly exceeding the previous experimental observation, providing a platform for efficiently generating and controlling spin current. Additionally, the platform is also anticipated to exhibit a considerable and tunable inverse effect, namely the spin-to-charge effect~\cite{baiEfficientSpintoChargeConversion2023}. This opens a promising avenue for future spintronic devices and memory storage applications.

\begin{acknowledgments}
\textit{Acknowledgments.}---We thank Yongpan Li for the helpful discussion. The work is supported by the National Key R\&D Program of China (Grant No. 2020YFA0308800), the NSF of China (Grant No. 12374055), and the Science Fund for Creative Research Groups of NSFC (Grant No. 12321004).
\end{acknowledgments}

\bibliography{ref}

\begin{thebibliography}{68}%
\makeatletter
\providecommand \@ifxundefined [1]{%
 \@ifx{#1\undefined}
}%
\providecommand \@ifnum [1]{%
 \ifnum #1\expandafter \@firstoftwo
 \else \expandafter \@secondoftwo
 \fi
}%
\providecommand \@ifx [1]{%
 \ifx #1\expandafter \@firstoftwo
 \else \expandafter \@secondoftwo
 \fi
}%
\providecommand \natexlab [1]{#1}%
\providecommand \enquote  [1]{``#1''}%
\providecommand \bibnamefont  [1]{#1}%
\providecommand \bibfnamefont [1]{#1}%
\providecommand \citenamefont [1]{#1}%
\providecommand \href@noop [0]{\@secondoftwo}%
\providecommand \href [0]{\begingroup \@sanitize@url \@href}%
\providecommand \@href[1]{\@@startlink{#1}\@@href}%
\providecommand \@@href[1]{\endgroup#1\@@endlink}%
\providecommand \@sanitize@url [0]{\catcode `\\12\catcode `\$12\catcode
  `\&12\catcode `\#12\catcode `\^12\catcode `\_12\catcode `\%12\relax}%
\providecommand \@@startlink[1]{}%
\providecommand \@@endlink[0]{}%
\providecommand \url  [0]{\begingroup\@sanitize@url \@url }%
\providecommand \@url [1]{\endgroup\@href {#1}{\urlprefix }}%
\providecommand \urlprefix  [0]{URL }%
\providecommand \Eprint [0]{\href }%
\providecommand \doibase [0]{http://dx.doi.org/}%
\providecommand \selectlanguage [0]{\@gobble}%
\providecommand \bibinfo  [0]{\@secondoftwo}%
\providecommand \bibfield  [0]{\@secondoftwo}%
\providecommand \translation [1]{[#1]}%
\providecommand \BibitemOpen [0]{}%
\providecommand \bibitemStop [0]{}%
\providecommand \bibitemNoStop [0]{.\EOS\space}%
\providecommand \EOS [0]{\spacefactor3000\relax}%
\providecommand \BibitemShut  [1]{\csname bibitem#1\endcsname}%
\let\auto@bib@innerbib\@empty
\bibitem [{\citenamefont {{\v S}mejkal}\ \emph
  {et~al.}(2022{\natexlab{a}})\citenamefont {{\v S}mejkal}, \citenamefont
  {Sinova},\ and\ \citenamefont
  {Jungwirth}}]{smejkalConventionalFerromagnetismAntiferromagnetism2022}%
  \BibitemOpen
  \bibfield  {author} {\bibinfo {author} {\bibfnamefont {L.}~\bibnamefont {{\v
  S}mejkal}}, \bibinfo {author} {\bibfnamefont {J.}~\bibnamefont {Sinova}}, \
  and\ \bibinfo {author} {\bibfnamefont {T.}~\bibnamefont {Jungwirth}},\ }\href
  {\doibase 10.1103/PhysRevX.12.031042} {\bibfield  {journal} {\bibinfo
  {journal} {Phys. Rev. X}\ }\textbf {\bibinfo {volume} {12}},\ \bibinfo
  {pages} {031042} (\bibinfo {year} {2022}{\natexlab{a}})}\BibitemShut
  {NoStop}%
\bibitem [{\citenamefont {{\v S}mejkal}\ \emph
  {et~al.}(2022{\natexlab{b}})\citenamefont {{\v S}mejkal}, \citenamefont
  {Sinova},\ and\ \citenamefont
  {Jungwirth}}]{smejkalEmergingResearchLandscape2022}%
  \BibitemOpen
  \bibfield  {author} {\bibinfo {author} {\bibfnamefont {L.}~\bibnamefont {{\v
  S}mejkal}}, \bibinfo {author} {\bibfnamefont {J.}~\bibnamefont {Sinova}}, \
  and\ \bibinfo {author} {\bibfnamefont {T.}~\bibnamefont {Jungwirth}},\ }\href
  {\doibase 10.1103/PhysRevX.12.040501} {\bibfield  {journal} {\bibinfo
  {journal} {Phys. Rev. X}\ }\textbf {\bibinfo {volume} {12}},\ \bibinfo
  {pages} {040501} (\bibinfo {year} {2022}{\natexlab{b}})}\BibitemShut
  {NoStop}%
\bibitem [{\citenamefont {Feng}\ \emph {et~al.}(2022)\citenamefont {Feng},
  \citenamefont {Zhou}, \citenamefont {{\v S}mejkal}, \citenamefont {Wu},
  \citenamefont {Zhu}, \citenamefont {Guo}, \citenamefont
  {{Gonz{\'a}lez-Hern{\'a}ndez}}, \citenamefont {Wang}, \citenamefont {Yan},
  \citenamefont {Qin}, \citenamefont {Zhang}, \citenamefont {Wu}, \citenamefont
  {Chen}, \citenamefont {Meng}, \citenamefont {Liu}, \citenamefont {Xia},
  \citenamefont {Sinova}, \citenamefont {Jungwirth},\ and\ \citenamefont
  {Liu}}]{fengAnomalousHallEffect2022}%
  \BibitemOpen
  \bibfield  {author} {\bibinfo {author} {\bibfnamefont {Z.}~\bibnamefont
  {Feng}}, \bibinfo {author} {\bibfnamefont {X.}~\bibnamefont {Zhou}}, \bibinfo
  {author} {\bibfnamefont {L.}~\bibnamefont {{\v S}mejkal}}, \bibinfo {author}
  {\bibfnamefont {L.}~\bibnamefont {Wu}}, \bibinfo {author} {\bibfnamefont
  {Z.}~\bibnamefont {Zhu}}, \bibinfo {author} {\bibfnamefont {H.}~\bibnamefont
  {Guo}}, \bibinfo {author} {\bibfnamefont {R.}~\bibnamefont
  {{Gonz{\'a}lez-Hern{\'a}ndez}}}, \bibinfo {author} {\bibfnamefont
  {X.}~\bibnamefont {Wang}}, \bibinfo {author} {\bibfnamefont {H.}~\bibnamefont
  {Yan}}, \bibinfo {author} {\bibfnamefont {P.}~\bibnamefont {Qin}}, \bibinfo
  {author} {\bibfnamefont {X.}~\bibnamefont {Zhang}}, \bibinfo {author}
  {\bibfnamefont {H.}~\bibnamefont {Wu}}, \bibinfo {author} {\bibfnamefont
  {H.}~\bibnamefont {Chen}}, \bibinfo {author} {\bibfnamefont {Z.}~\bibnamefont
  {Meng}}, \bibinfo {author} {\bibfnamefont {L.}~\bibnamefont {Liu}}, \bibinfo
  {author} {\bibfnamefont {Z.}~\bibnamefont {Xia}}, \bibinfo {author}
  {\bibfnamefont {J.}~\bibnamefont {Sinova}}, \bibinfo {author} {\bibfnamefont
  {T.}~\bibnamefont {Jungwirth}}, \ and\ \bibinfo {author} {\bibfnamefont
  {Z.}~\bibnamefont {Liu}},\ }\href {\doibase 10.1038/s41928-022-00866-z}
  {\bibfield  {journal} {\bibinfo  {journal} {Nat Electron}\ }\textbf {\bibinfo
  {volume} {5}},\ \bibinfo {pages} {735} (\bibinfo {year} {2022})}\BibitemShut
  {NoStop}%
\bibitem [{\citenamefont {Bai}\ \emph {et~al.}(2023)\citenamefont {Bai},
  \citenamefont {Zhang}, \citenamefont {Zhou}, \citenamefont {Chen},
  \citenamefont {Wan}, \citenamefont {Han}, \citenamefont {Zhu}, \citenamefont
  {Liang}, \citenamefont {Su}, \citenamefont {Han}, \citenamefont {Pan},\ and\
  \citenamefont {Song}}]{baiEfficientSpintoChargeConversion2023}%
  \BibitemOpen
  \bibfield  {author} {\bibinfo {author} {\bibfnamefont {H.}~\bibnamefont
  {Bai}}, \bibinfo {author} {\bibfnamefont {Y.~C.}\ \bibnamefont {Zhang}},
  \bibinfo {author} {\bibfnamefont {Y.~J.}\ \bibnamefont {Zhou}}, \bibinfo
  {author} {\bibfnamefont {P.}~\bibnamefont {Chen}}, \bibinfo {author}
  {\bibfnamefont {C.~H.}\ \bibnamefont {Wan}}, \bibinfo {author} {\bibfnamefont
  {L.}~\bibnamefont {Han}}, \bibinfo {author} {\bibfnamefont {W.~X.}\
  \bibnamefont {Zhu}}, \bibinfo {author} {\bibfnamefont {S.~X.}\ \bibnamefont
  {Liang}}, \bibinfo {author} {\bibfnamefont {Y.~C.}\ \bibnamefont {Su}},
  \bibinfo {author} {\bibfnamefont {X.~F.}\ \bibnamefont {Han}}, \bibinfo
  {author} {\bibfnamefont {F.}~\bibnamefont {Pan}}, \ and\ \bibinfo {author}
  {\bibfnamefont {C.}~\bibnamefont {Song}},\ }\href {\doibase
  10.1103/PhysRevLett.130.216701} {\bibfield  {journal} {\bibinfo  {journal}
  {Phys. Rev. Lett.}\ }\textbf {\bibinfo {volume} {130}},\ \bibinfo {pages}
  {216701} (\bibinfo {year} {2023})}\BibitemShut {NoStop}%
\bibitem [{\citenamefont {Zhou}\ \emph {et~al.}(2024)\citenamefont {Zhou},
  \citenamefont {Feng}, \citenamefont {Zhang}, \citenamefont {{\v S}mejkal},
  \citenamefont {Sinova}, \citenamefont {Mokrousov},\ and\ \citenamefont
  {Yao}}]{zhouCrystalThermalTransport2024a}%
  \BibitemOpen
  \bibfield  {author} {\bibinfo {author} {\bibfnamefont {X.}~\bibnamefont
  {Zhou}}, \bibinfo {author} {\bibfnamefont {W.}~\bibnamefont {Feng}}, \bibinfo
  {author} {\bibfnamefont {R.-W.}\ \bibnamefont {Zhang}}, \bibinfo {author}
  {\bibfnamefont {L.}~\bibnamefont {{\v S}mejkal}}, \bibinfo {author}
  {\bibfnamefont {J.}~\bibnamefont {Sinova}}, \bibinfo {author} {\bibfnamefont
  {Y.}~\bibnamefont {Mokrousov}}, \ and\ \bibinfo {author} {\bibfnamefont
  {Y.}~\bibnamefont {Yao}},\ }\href {\doibase 10.1103/PhysRevLett.132.056701}
  {\bibfield  {journal} {\bibinfo  {journal} {Phys. Rev. Lett.}\ }\textbf
  {\bibinfo {volume} {132}},\ \bibinfo {pages} {056701} (\bibinfo {year}
  {2024})}\BibitemShut {NoStop}%
\bibitem [{\citenamefont {Ouassou}\ \emph {et~al.}(2023)\citenamefont
  {Ouassou}, \citenamefont {Brataas},\ and\ \citenamefont
  {Linder}}]{ouassouDcJosephsonEffect2023}%
  \BibitemOpen
  \bibfield  {author} {\bibinfo {author} {\bibfnamefont {J.~A.}\ \bibnamefont
  {Ouassou}}, \bibinfo {author} {\bibfnamefont {A.}~\bibnamefont {Brataas}}, \
  and\ \bibinfo {author} {\bibfnamefont {J.}~\bibnamefont {Linder}},\ }\href
  {\doibase 10.1103/PhysRevLett.131.076003} {\bibfield  {journal} {\bibinfo
  {journal} {Phys. Rev. Lett.}\ }\textbf {\bibinfo {volume} {131}},\ \bibinfo
  {pages} {076003} (\bibinfo {year} {2023})}\BibitemShut {NoStop}%
\bibitem [{\citenamefont {Li}\ and\ \citenamefont
  {Liu}(2023)}]{liMajoranaCornerModes2023}%
  \BibitemOpen
  \bibfield  {author} {\bibinfo {author} {\bibfnamefont {Y.-X.}\ \bibnamefont
  {Li}}\ and\ \bibinfo {author} {\bibfnamefont {C.-C.}\ \bibnamefont {Liu}},\
  }\href {\doibase 10.1103/PhysRevB.108.205410} {\bibfield  {journal} {\bibinfo
   {journal} {Phys. Rev. B}\ }\textbf {\bibinfo {volume} {108}},\ \bibinfo
  {pages} {205410} (\bibinfo {year} {2023})}\BibitemShut {NoStop}%
\bibitem [{\citenamefont {Zhu}\ \emph {et~al.}(2023)\citenamefont {Zhu},
  \citenamefont {Zhuang}, \citenamefont {Wu},\ and\ \citenamefont
  {Yan}}]{zhuTopologicalSuperconductivityTwodimensional2023}%
  \BibitemOpen
  \bibfield  {author} {\bibinfo {author} {\bibfnamefont {D.}~\bibnamefont
  {Zhu}}, \bibinfo {author} {\bibfnamefont {Z.-Y.}\ \bibnamefont {Zhuang}},
  \bibinfo {author} {\bibfnamefont {Z.}~\bibnamefont {Wu}}, \ and\ \bibinfo
  {author} {\bibfnamefont {Z.}~\bibnamefont {Yan}},\ }\href {\doibase
  10.1103/PhysRevB.108.184505} {\bibfield  {journal} {\bibinfo  {journal}
  {Phys. Rev. B}\ }\textbf {\bibinfo {volume} {108}},\ \bibinfo {pages}
  {184505} (\bibinfo {year} {2023})}\BibitemShut {NoStop}%
\bibitem [{\citenamefont {Zhang}\ \emph {et~al.}(2024)\citenamefont {Zhang},
  \citenamefont {Hu},\ and\ \citenamefont
  {Neupert}}]{zhangFinitemomentumCooperPairing2024}%
  \BibitemOpen
  \bibfield  {author} {\bibinfo {author} {\bibfnamefont {S.-B.}\ \bibnamefont
  {Zhang}}, \bibinfo {author} {\bibfnamefont {L.-H.}\ \bibnamefont {Hu}}, \
  and\ \bibinfo {author} {\bibfnamefont {T.}~\bibnamefont {Neupert}},\ }\href
  {\doibase 10.1038/s41467-024-45951-3} {\bibfield  {journal} {\bibinfo
  {journal} {Nat Commun}\ }\textbf {\bibinfo {volume} {15}},\ \bibinfo {pages}
  {1801} (\bibinfo {year} {2024})}\BibitemShut {NoStop}%
\bibitem [{\citenamefont {{\v S}mejkal}\ \emph {et~al.}(2020)\citenamefont {{\v
  S}mejkal}, \citenamefont {{Gonz{\'a}lez-Hern{\'a}ndez}}, \citenamefont
  {Jungwirth},\ and\ \citenamefont
  {Sinova}}]{smejkalCrystalTimereversalSymmetry2020a}%
  \BibitemOpen
  \bibfield  {author} {\bibinfo {author} {\bibfnamefont {L.}~\bibnamefont {{\v
  S}mejkal}}, \bibinfo {author} {\bibfnamefont {R.}~\bibnamefont
  {{Gonz{\'a}lez-Hern{\'a}ndez}}}, \bibinfo {author} {\bibfnamefont
  {T.}~\bibnamefont {Jungwirth}}, \ and\ \bibinfo {author} {\bibfnamefont
  {J.}~\bibnamefont {Sinova}},\ }\href {\doibase 10.1126/sciadv.aaz8809}
  {\bibfield  {journal} {\bibinfo  {journal} {Sci. Adv}\ }\textbf {\bibinfo
  {volume} {6}},\ \bibinfo {pages} {eaaz8809} (\bibinfo {year}
  {2020})}\BibitemShut {NoStop}%
\bibitem [{\citenamefont {Lee}\ \emph {et~al.}(2024)\citenamefont {Lee},
  \citenamefont {Lee}, \citenamefont {Jung}, \citenamefont {Jung},
  \citenamefont {Kim}, \citenamefont {Lee}, \citenamefont {Seok}, \citenamefont
  {Kim}, \citenamefont {Park}, \citenamefont {{\v S}mejkal}, \citenamefont
  {Kang},\ and\ \citenamefont {Kim}}]{leeBrokenKramersDegeneracy2024a}%
  \BibitemOpen
  \bibfield  {author} {\bibinfo {author} {\bibfnamefont {S.}~\bibnamefont
  {Lee}}, \bibinfo {author} {\bibfnamefont {S.}~\bibnamefont {Lee}}, \bibinfo
  {author} {\bibfnamefont {S.}~\bibnamefont {Jung}}, \bibinfo {author}
  {\bibfnamefont {J.}~\bibnamefont {Jung}}, \bibinfo {author} {\bibfnamefont
  {D.}~\bibnamefont {Kim}}, \bibinfo {author} {\bibfnamefont {Y.}~\bibnamefont
  {Lee}}, \bibinfo {author} {\bibfnamefont {B.}~\bibnamefont {Seok}}, \bibinfo
  {author} {\bibfnamefont {J.}~\bibnamefont {Kim}}, \bibinfo {author}
  {\bibfnamefont {B.~G.}\ \bibnamefont {Park}}, \bibinfo {author}
  {\bibfnamefont {L.}~\bibnamefont {{\v S}mejkal}}, \bibinfo {author}
  {\bibfnamefont {C.-J.}\ \bibnamefont {Kang}}, \ and\ \bibinfo {author}
  {\bibfnamefont {C.}~\bibnamefont {Kim}},\ }\href {\doibase
  10.1103/PhysRevLett.132.036702} {\bibfield  {journal} {\bibinfo  {journal}
  {Phys. Rev. Lett.}\ }\textbf {\bibinfo {volume} {132}},\ \bibinfo {pages}
  {036702} (\bibinfo {year} {2024})}\BibitemShut {NoStop}%
\bibitem [{\citenamefont {Mazin}(2023)}]{mazinAltermagnetismMnTeOrigin2023a}%
  \BibitemOpen
  \bibfield  {author} {\bibinfo {author} {\bibfnamefont {I.~I.}\ \bibnamefont
  {Mazin}},\ }\href {\doibase 10.1103/PhysRevB.107.L100418} {\bibfield
  {journal} {\bibinfo  {journal} {Phys. Rev. B}\ }\textbf {\bibinfo {volume}
  {107}},\ \bibinfo {pages} {L100418} (\bibinfo {year} {2023})}\BibitemShut
  {NoStop}%
\bibitem [{\citenamefont {Kriegner}\ \emph {et~al.}(2017)\citenamefont
  {Kriegner}, \citenamefont {Reichlova}, \citenamefont {Grenzer}, \citenamefont
  {Schmidt}, \citenamefont {Ressouche}, \citenamefont {Godinho}, \citenamefont
  {Wagner}, \citenamefont {Martin}, \citenamefont {Shick}, \citenamefont
  {Volobuev}, \citenamefont {Springholz}, \citenamefont {Hol{\'y}},
  \citenamefont {Wunderlich}, \citenamefont {Jungwirth},\ and\ \citenamefont
  {V{\'y}born{\'y}}}]{kriegnerMagneticAnisotropyAntiferromagnetic2017}%
  \BibitemOpen
  \bibfield  {author} {\bibinfo {author} {\bibfnamefont {D.}~\bibnamefont
  {Kriegner}}, \bibinfo {author} {\bibfnamefont {H.}~\bibnamefont {Reichlova}},
  \bibinfo {author} {\bibfnamefont {J.}~\bibnamefont {Grenzer}}, \bibinfo
  {author} {\bibfnamefont {W.}~\bibnamefont {Schmidt}}, \bibinfo {author}
  {\bibfnamefont {E.}~\bibnamefont {Ressouche}}, \bibinfo {author}
  {\bibfnamefont {J.}~\bibnamefont {Godinho}}, \bibinfo {author} {\bibfnamefont
  {T.}~\bibnamefont {Wagner}}, \bibinfo {author} {\bibfnamefont {S.~Y.}\
  \bibnamefont {Martin}}, \bibinfo {author} {\bibfnamefont {A.~B.}\
  \bibnamefont {Shick}}, \bibinfo {author} {\bibfnamefont {V.~V.}\ \bibnamefont
  {Volobuev}}, \bibinfo {author} {\bibfnamefont {G.}~\bibnamefont
  {Springholz}}, \bibinfo {author} {\bibfnamefont {V.}~\bibnamefont
  {Hol{\'y}}}, \bibinfo {author} {\bibfnamefont {J.}~\bibnamefont
  {Wunderlich}}, \bibinfo {author} {\bibfnamefont {T.}~\bibnamefont
  {Jungwirth}}, \ and\ \bibinfo {author} {\bibfnamefont {K.}~\bibnamefont
  {V{\'y}born{\'y}}},\ }\href {\doibase 10.1103/PhysRevB.96.214418} {\bibfield
  {journal} {\bibinfo  {journal} {Phys. Rev. B}\ }\textbf {\bibinfo {volume}
  {96}},\ \bibinfo {pages} {214418} (\bibinfo {year} {2017})}\BibitemShut
  {NoStop}%
\bibitem [{\citenamefont {Lovesey}\ \emph {et~al.}(2023)\citenamefont
  {Lovesey}, \citenamefont {Khalyavin},\ and\ \citenamefont {{van der
  Laan}}}]{loveseyTemplatesMagneticSymmetry2023}%
  \BibitemOpen
  \bibfield  {author} {\bibinfo {author} {\bibfnamefont {S.~W.}\ \bibnamefont
  {Lovesey}}, \bibinfo {author} {\bibfnamefont {D.~D.}\ \bibnamefont
  {Khalyavin}}, \ and\ \bibinfo {author} {\bibfnamefont {G.}~\bibnamefont {{van
  der Laan}}},\ }\href {\doibase 10.1103/PhysRevB.108.174437} {\bibfield
  {journal} {\bibinfo  {journal} {Phys. Rev. B}\ }\textbf {\bibinfo {volume}
  {108}},\ \bibinfo {pages} {174437} (\bibinfo {year} {2023})}\BibitemShut
  {NoStop}%
\bibitem [{\citenamefont {Yuan}\ \emph {et~al.}(2020)\citenamefont {Yuan},
  \citenamefont {Wang}, \citenamefont {Luo}, \citenamefont {Rashba},\ and\
  \citenamefont {Zunger}}]{yuanGiantMomentumdependentSpin2020a}%
  \BibitemOpen
  \bibfield  {author} {\bibinfo {author} {\bibfnamefont {L.-D.}\ \bibnamefont
  {Yuan}}, \bibinfo {author} {\bibfnamefont {Z.}~\bibnamefont {Wang}}, \bibinfo
  {author} {\bibfnamefont {J.-W.}\ \bibnamefont {Luo}}, \bibinfo {author}
  {\bibfnamefont {E.~I.}\ \bibnamefont {Rashba}}, \ and\ \bibinfo {author}
  {\bibfnamefont {A.}~\bibnamefont {Zunger}},\ }\href {\doibase
  10.1103/PhysRevB.102.014422} {\bibfield  {journal} {\bibinfo  {journal}
  {Phys. Rev. B}\ }\textbf {\bibinfo {volume} {102}},\ \bibinfo {pages}
  {014422} (\bibinfo {year} {2020})}\BibitemShut {NoStop}%
\bibitem [{\citenamefont {Liu}\ \emph {et~al.}()\citenamefont {Liu},
  \citenamefont {Kang}, \citenamefont {Wang}, \citenamefont {Gao},
  \citenamefont {Qi}, \citenamefont {Zhao},\ and\ \citenamefont
  {Jiang}}]{liuInverseMagnetocaloricEffect}%
  \BibitemOpen
  \bibfield  {author} {\bibinfo {author} {\bibfnamefont {Q.}~\bibnamefont
  {Liu}}, \bibinfo {author} {\bibfnamefont {J.}~\bibnamefont {Kang}}, \bibinfo
  {author} {\bibfnamefont {P.}~\bibnamefont {Wang}}, \bibinfo {author}
  {\bibfnamefont {W.}~\bibnamefont {Gao}}, \bibinfo {author} {\bibfnamefont
  {Y.}~\bibnamefont {Qi}}, \bibinfo {author} {\bibfnamefont {J.}~\bibnamefont
  {Zhao}}, \ and\ \bibinfo {author} {\bibfnamefont {X.}~\bibnamefont {Jiang}},\
  }\href {\doibase 10.1002/adfm.202402080} {\bibfield  {journal} {\bibinfo
  {journal} {Adv. Funct}\ }\textbf {\bibinfo {volume} {n/a}},\ \bibinfo {pages}
  {2402080}}\BibitemShut {NoStop}%
\bibitem [{\citenamefont {Brekke}\ \emph {et~al.}(2023)\citenamefont {Brekke},
  \citenamefont {Brataas},\ and\ \citenamefont
  {Sudb{\o}}}]{brekkeTwodimensionalAltermagnetsMinimal2023}%
  \BibitemOpen
  \bibfield  {author} {\bibinfo {author} {\bibfnamefont {B.}~\bibnamefont
  {Brekke}}, \bibinfo {author} {\bibfnamefont {A.}~\bibnamefont {Brataas}}, \
  and\ \bibinfo {author} {\bibfnamefont {A.}~\bibnamefont {Sudb{\o}}},\ }\href
  {\doibase 10.1103/PhysRevB.108.224421} {\bibfield  {journal} {\bibinfo
  {journal} {Phys. Rev. B}\ }\textbf {\bibinfo {volume} {108}},\ \bibinfo
  {pages} {224421} (\bibinfo {year} {2023})}\BibitemShut {NoStop}%
\bibitem [{\citenamefont {S{\o}dequist}\ and\ \citenamefont
  {Olsen}(2024)}]{sodequistTwodimensionalAltermagnetsHigh2024}%
  \BibitemOpen
  \bibfield  {author} {\bibinfo {author} {\bibfnamefont {J.}~\bibnamefont
  {S{\o}dequist}}\ and\ \bibinfo {author} {\bibfnamefont {T.}~\bibnamefont
  {Olsen}},\ }\href {\doibase 10.48550/arXiv.2401.05992} {\bibfield  {journal}
  {\bibinfo  {journal} {arXiv:2401.05992}\ } (\bibinfo {year} {2024}),\
  10.48550/arXiv.2401.05992},\ \Eprint {http://arxiv.org/abs/2401.05992}
  {2401.05992 [cond-mat]} \BibitemShut {NoStop}%
\bibitem [{\citenamefont {Mazin}\ \emph {et~al.}(2023)\citenamefont {Mazin},
  \citenamefont {{Gonz{\'a}lez-Hern{\'a}ndez}},\ and\ \citenamefont {{\v
  S}mejkal}}]{mazinInducedMonolayerAltermagnetism2023}%
  \BibitemOpen
  \bibfield  {author} {\bibinfo {author} {\bibfnamefont {I.}~\bibnamefont
  {Mazin}}, \bibinfo {author} {\bibfnamefont {R.}~\bibnamefont
  {{Gonz{\'a}lez-Hern{\'a}ndez}}}, \ and\ \bibinfo {author} {\bibfnamefont
  {L.}~\bibnamefont {{\v S}mejkal}},\ }\href@noop {} {\bibfield  {journal}
  {\bibinfo  {journal} {arXiv:2309.02355}\ } (\bibinfo {year}
  {2023})}\BibitemShut {NoStop}%
\bibitem [{\citenamefont {Guo}\ \emph {et~al.}(2023)\citenamefont {Guo},
  \citenamefont {Guo}, \citenamefont {Cheng}, \citenamefont {Wang},\ and\
  \citenamefont {Ang}}]{guoPiezoelectricAltermagnetismSpinvalley2023}%
  \BibitemOpen
  \bibfield  {author} {\bibinfo {author} {\bibfnamefont {S.-D.}\ \bibnamefont
  {Guo}}, \bibinfo {author} {\bibfnamefont {X.-S.}\ \bibnamefont {Guo}},
  \bibinfo {author} {\bibfnamefont {K.}~\bibnamefont {Cheng}}, \bibinfo
  {author} {\bibfnamefont {K.}~\bibnamefont {Wang}}, \ and\ \bibinfo {author}
  {\bibfnamefont {Y.~S.}\ \bibnamefont {Ang}},\ }\href {\doibase
  10.1063/5.0161431} {\bibfield  {journal} {\bibinfo  {journal} {Appl. Phys.
  Lett}\ }\textbf {\bibinfo {volume} {123}},\ \bibinfo {pages} {082401}
  (\bibinfo {year} {2023})}\BibitemShut {NoStop}%
\bibitem [{\citenamefont {Huang}\ \emph {et~al.}(2017)\citenamefont {Huang},
  \citenamefont {Clark}, \citenamefont {{Navarro-Moratalla}}, \citenamefont
  {Klein}, \citenamefont {Cheng}, \citenamefont {Seyler}, \citenamefont
  {Zhong}, \citenamefont {Schmidgall}, \citenamefont {McGuire}, \citenamefont
  {Cobden}, \citenamefont {Yao}, \citenamefont {Xiao}, \citenamefont
  {{Jarillo-Herrero}},\ and\ \citenamefont
  {Xu}}]{huangLayerdependentFerromagnetismVan2017}%
  \BibitemOpen
  \bibfield  {author} {\bibinfo {author} {\bibfnamefont {B.}~\bibnamefont
  {Huang}}, \bibinfo {author} {\bibfnamefont {G.}~\bibnamefont {Clark}},
  \bibinfo {author} {\bibfnamefont {E.}~\bibnamefont {{Navarro-Moratalla}}},
  \bibinfo {author} {\bibfnamefont {D.~R.}\ \bibnamefont {Klein}}, \bibinfo
  {author} {\bibfnamefont {R.}~\bibnamefont {Cheng}}, \bibinfo {author}
  {\bibfnamefont {K.~L.}\ \bibnamefont {Seyler}}, \bibinfo {author}
  {\bibfnamefont {D.}~\bibnamefont {Zhong}}, \bibinfo {author} {\bibfnamefont
  {E.}~\bibnamefont {Schmidgall}}, \bibinfo {author} {\bibfnamefont {M.~A.}\
  \bibnamefont {McGuire}}, \bibinfo {author} {\bibfnamefont {D.~H.}\
  \bibnamefont {Cobden}}, \bibinfo {author} {\bibfnamefont {W.}~\bibnamefont
  {Yao}}, \bibinfo {author} {\bibfnamefont {D.}~\bibnamefont {Xiao}}, \bibinfo
  {author} {\bibfnamefont {P.}~\bibnamefont {{Jarillo-Herrero}}}, \ and\
  \bibinfo {author} {\bibfnamefont {X.}~\bibnamefont {Xu}},\ }\href {\doibase
  10.1038/nature22391} {\bibfield  {journal} {\bibinfo  {journal} {Nature}\
  }\textbf {\bibinfo {volume} {546}},\ \bibinfo {pages} {270} (\bibinfo {year}
  {2017})}\BibitemShut {NoStop}%
\bibitem [{\citenamefont {{Bedoya-Pinto}}\ \emph {et~al.}(2021)\citenamefont
  {{Bedoya-Pinto}}, \citenamefont {Ji}, \citenamefont {Pandeya}, \citenamefont
  {Gargiani}, \citenamefont {Valvidares}, \citenamefont {Sessi}, \citenamefont
  {Taylor}, \citenamefont {Radu}, \citenamefont {Chang},\ and\ \citenamefont
  {Parkin}}]{bedoya-pintoIntrinsic2DXYFerromagnetism2021}%
  \BibitemOpen
  \bibfield  {author} {\bibinfo {author} {\bibfnamefont {A.}~\bibnamefont
  {{Bedoya-Pinto}}}, \bibinfo {author} {\bibfnamefont {J.-R.}\ \bibnamefont
  {Ji}}, \bibinfo {author} {\bibfnamefont {A.~K.}\ \bibnamefont {Pandeya}},
  \bibinfo {author} {\bibfnamefont {P.}~\bibnamefont {Gargiani}}, \bibinfo
  {author} {\bibfnamefont {M.}~\bibnamefont {Valvidares}}, \bibinfo {author}
  {\bibfnamefont {P.}~\bibnamefont {Sessi}}, \bibinfo {author} {\bibfnamefont
  {J.~M.}\ \bibnamefont {Taylor}}, \bibinfo {author} {\bibfnamefont
  {F.}~\bibnamefont {Radu}}, \bibinfo {author} {\bibfnamefont {K.}~\bibnamefont
  {Chang}}, \ and\ \bibinfo {author} {\bibfnamefont {S.~S.~P.}\ \bibnamefont
  {Parkin}},\ }\href {\doibase 10.1126/science.abd5146} {\bibfield  {journal}
  {\bibinfo  {journal} {Science}\ }\textbf {\bibinfo {volume} {374}},\ \bibinfo
  {pages} {616} (\bibinfo {year} {2021})}\BibitemShut {NoStop}%
\bibitem [{\citenamefont {Zhang}\ \emph
  {et~al.}(2019{\natexlab{a}})\citenamefont {Zhang}, \citenamefont {Shang},
  \citenamefont {Jiang}, \citenamefont {Rasmita}, \citenamefont {Gao},\ and\
  \citenamefont {Yu}}]{zhangDirectPhotoluminescenceProbing2019}%
  \BibitemOpen
  \bibfield  {author} {\bibinfo {author} {\bibfnamefont {Z.}~\bibnamefont
  {Zhang}}, \bibinfo {author} {\bibfnamefont {J.}~\bibnamefont {Shang}},
  \bibinfo {author} {\bibfnamefont {C.}~\bibnamefont {Jiang}}, \bibinfo
  {author} {\bibfnamefont {A.}~\bibnamefont {Rasmita}}, \bibinfo {author}
  {\bibfnamefont {W.}~\bibnamefont {Gao}}, \ and\ \bibinfo {author}
  {\bibfnamefont {T.}~\bibnamefont {Yu}},\ }\href {\doibase
  10.1021/acs.nanolett.9b00553} {\bibfield  {journal} {\bibinfo  {journal}
  {Nano Lett.}\ }\textbf {\bibinfo {volume} {19}},\ \bibinfo {pages} {3138}
  (\bibinfo {year} {2019}{\natexlab{a}})}\BibitemShut {NoStop}%
\bibitem [{\citenamefont {Lee}\ \emph {et~al.}(2021)\citenamefont {Lee},
  \citenamefont {Dismukes}, \citenamefont {Telford}, \citenamefont {Wiscons},
  \citenamefont {Wang}, \citenamefont {Xu}, \citenamefont {Nuckolls},
  \citenamefont {Dean}, \citenamefont {Roy},\ and\ \citenamefont
  {Zhu}}]{leeMagneticOrderSymmetry2021}%
  \BibitemOpen
  \bibfield  {author} {\bibinfo {author} {\bibfnamefont {K.}~\bibnamefont
  {Lee}}, \bibinfo {author} {\bibfnamefont {A.~H.}\ \bibnamefont {Dismukes}},
  \bibinfo {author} {\bibfnamefont {E.~J.}\ \bibnamefont {Telford}}, \bibinfo
  {author} {\bibfnamefont {R.~A.}\ \bibnamefont {Wiscons}}, \bibinfo {author}
  {\bibfnamefont {J.}~\bibnamefont {Wang}}, \bibinfo {author} {\bibfnamefont
  {X.}~\bibnamefont {Xu}}, \bibinfo {author} {\bibfnamefont {C.}~\bibnamefont
  {Nuckolls}}, \bibinfo {author} {\bibfnamefont {C.~R.}\ \bibnamefont {Dean}},
  \bibinfo {author} {\bibfnamefont {X.}~\bibnamefont {Roy}}, \ and\ \bibinfo
  {author} {\bibfnamefont {X.}~\bibnamefont {Zhu}},\ }\href {\doibase
  10.1021/acs.nanolett.1c00219} {\bibfield  {journal} {\bibinfo  {journal}
  {Nano Lett.}\ }\textbf {\bibinfo {volume} {21}},\ \bibinfo {pages} {3511}
  (\bibinfo {year} {2021})}\BibitemShut {NoStop}%
\bibitem [{\citenamefont {Aapro}\ \emph {et~al.}(2021)\citenamefont {Aapro},
  \citenamefont {Huda}, \citenamefont {Karthikeyan}, \citenamefont
  {Kezilebieke}, \citenamefont {Ganguli}, \citenamefont {Herrero},
  \citenamefont {Huang}, \citenamefont {Liljeroth},\ and\ \citenamefont
  {Komsa}}]{aaproSynthesisPropertiesMonolayer2021}%
  \BibitemOpen
  \bibfield  {author} {\bibinfo {author} {\bibfnamefont {M.}~\bibnamefont
  {Aapro}}, \bibinfo {author} {\bibfnamefont {M.~N.}\ \bibnamefont {Huda}},
  \bibinfo {author} {\bibfnamefont {J.}~\bibnamefont {Karthikeyan}}, \bibinfo
  {author} {\bibfnamefont {S.}~\bibnamefont {Kezilebieke}}, \bibinfo {author}
  {\bibfnamefont {S.~C.}\ \bibnamefont {Ganguli}}, \bibinfo {author}
  {\bibfnamefont {H.~G.}\ \bibnamefont {Herrero}}, \bibinfo {author}
  {\bibfnamefont {X.}~\bibnamefont {Huang}}, \bibinfo {author} {\bibfnamefont
  {P.}~\bibnamefont {Liljeroth}}, \ and\ \bibinfo {author} {\bibfnamefont
  {H.-P.}\ \bibnamefont {Komsa}},\ }\href {\doibase 10.1021/acsnano.1c05532}
  {\bibfield  {journal} {\bibinfo  {journal} {ACS Nano}\ }\textbf {\bibinfo
  {volume} {15}},\ \bibinfo {pages} {13794} (\bibinfo {year}
  {2021})}\BibitemShut {NoStop}%
\bibitem [{\citenamefont {Otrokov}\ \emph {et~al.}(2019)\citenamefont
  {Otrokov}, \citenamefont {Klimovskikh}, \citenamefont {Bentmann},
  \citenamefont {Estyunin}, \citenamefont {Zeugner}, \citenamefont {Aliev},
  \citenamefont {Ga{\ss}}, \citenamefont {Wolter}, \citenamefont {Koroleva},
  \citenamefont {Shikin} \emph {et~al.}}]{MnBiTe2019_Nature}%
  \BibitemOpen
  \bibfield  {author} {\bibinfo {author} {\bibfnamefont {M.~M.}\ \bibnamefont
  {Otrokov}}, \bibinfo {author} {\bibfnamefont {I.~I.}\ \bibnamefont
  {Klimovskikh}}, \bibinfo {author} {\bibfnamefont {H.}~\bibnamefont
  {Bentmann}}, \bibinfo {author} {\bibfnamefont {D.}~\bibnamefont {Estyunin}},
  \bibinfo {author} {\bibfnamefont {A.}~\bibnamefont {Zeugner}}, \bibinfo
  {author} {\bibfnamefont {Z.~S.}\ \bibnamefont {Aliev}}, \bibinfo {author}
  {\bibfnamefont {S.}~\bibnamefont {Ga{\ss}}}, \bibinfo {author} {\bibfnamefont
  {A.}~\bibnamefont {Wolter}}, \bibinfo {author} {\bibfnamefont
  {A.}~\bibnamefont {Koroleva}}, \bibinfo {author} {\bibfnamefont {A.~M.}\
  \bibnamefont {Shikin}},  \emph {et~al.},\ }\href@noop {} {\bibfield
  {journal} {\bibinfo  {journal} {Nature}\ }\textbf {\bibinfo {volume} {576}},\
  \bibinfo {pages} {416} (\bibinfo {year} {2019})}\BibitemShut {NoStop}%
\bibitem [{\citenamefont {Zhang}\ \emph
  {et~al.}(2019{\natexlab{b}})\citenamefont {Zhang}, \citenamefont {Shi},
  \citenamefont {Zhu}, \citenamefont {Xing}, \citenamefont {Zhang},\ and\
  \citenamefont {Wang}}]{zhangTopologicalAxionStates2019prl}%
  \BibitemOpen
  \bibfield  {author} {\bibinfo {author} {\bibfnamefont {D.}~\bibnamefont
  {Zhang}}, \bibinfo {author} {\bibfnamefont {M.}~\bibnamefont {Shi}}, \bibinfo
  {author} {\bibfnamefont {T.}~\bibnamefont {Zhu}}, \bibinfo {author}
  {\bibfnamefont {D.}~\bibnamefont {Xing}}, \bibinfo {author} {\bibfnamefont
  {H.}~\bibnamefont {Zhang}}, \ and\ \bibinfo {author} {\bibfnamefont
  {J.}~\bibnamefont {Wang}},\ }\href {\doibase 10.1103/PhysRevLett.122.206401}
  {\bibfield  {journal} {\bibinfo  {journal} {Phys. Rev. Lett.}\ }\textbf
  {\bibinfo {volume} {122}},\ \bibinfo {pages} {206401} (\bibinfo {year}
  {2019}{\natexlab{b}})}\BibitemShut {NoStop}%
\bibitem [{\citenamefont {Li}\ \emph {et~al.}(2019)\citenamefont {Li},
  \citenamefont {Li}, \citenamefont {Du}, \citenamefont {Wang}, \citenamefont
  {Gu}, \citenamefont {Zhang}, \citenamefont {He}, \citenamefont {Duan},\ and\
  \citenamefont {Xu}}]{li2019intrinsicSA}%
  \BibitemOpen
  \bibfield  {author} {\bibinfo {author} {\bibfnamefont {J.}~\bibnamefont
  {Li}}, \bibinfo {author} {\bibfnamefont {Y.}~\bibnamefont {Li}}, \bibinfo
  {author} {\bibfnamefont {S.}~\bibnamefont {Du}}, \bibinfo {author}
  {\bibfnamefont {Z.}~\bibnamefont {Wang}}, \bibinfo {author} {\bibfnamefont
  {B.-L.}\ \bibnamefont {Gu}}, \bibinfo {author} {\bibfnamefont {S.-C.}\
  \bibnamefont {Zhang}}, \bibinfo {author} {\bibfnamefont {K.}~\bibnamefont
  {He}}, \bibinfo {author} {\bibfnamefont {W.}~\bibnamefont {Duan}}, \ and\
  \bibinfo {author} {\bibfnamefont {Y.}~\bibnamefont {Xu}},\ }\href@noop {}
  {\bibfield  {journal} {\bibinfo  {journal} {Science Advances}\ }\textbf
  {\bibinfo {volume} {5}},\ \bibinfo {pages} {eaaw5685} (\bibinfo {year}
  {2019})}\BibitemShut {NoStop}%
\bibitem [{\citenamefont {O'Hara}\ \emph {et~al.}(2018)\citenamefont {O'Hara},
  \citenamefont {Zhu}, \citenamefont {Trout}, \citenamefont {Ahmed},
  \citenamefont {Luo}, \citenamefont {Lee}, \citenamefont {Brenner},
  \citenamefont {Rajan}, \citenamefont {Gupta}, \citenamefont {McComb},\ and\
  \citenamefont {Kawakami}}]{oharaRoomTemperatureIntrinsic2018}%
  \BibitemOpen
  \bibfield  {author} {\bibinfo {author} {\bibfnamefont {D.~J.}\ \bibnamefont
  {O'Hara}}, \bibinfo {author} {\bibfnamefont {T.}~\bibnamefont {Zhu}},
  \bibinfo {author} {\bibfnamefont {A.~H.}\ \bibnamefont {Trout}}, \bibinfo
  {author} {\bibfnamefont {A.~S.}\ \bibnamefont {Ahmed}}, \bibinfo {author}
  {\bibfnamefont {Y.~K.}\ \bibnamefont {Luo}}, \bibinfo {author} {\bibfnamefont
  {C.~H.}\ \bibnamefont {Lee}}, \bibinfo {author} {\bibfnamefont {M.~R.}\
  \bibnamefont {Brenner}}, \bibinfo {author} {\bibfnamefont {S.}~\bibnamefont
  {Rajan}}, \bibinfo {author} {\bibfnamefont {J.~A.}\ \bibnamefont {Gupta}},
  \bibinfo {author} {\bibfnamefont {D.~W.}\ \bibnamefont {McComb}}, \ and\
  \bibinfo {author} {\bibfnamefont {R.~K.}\ \bibnamefont {Kawakami}},\ }\href
  {\doibase 10.1021/acs.nanolett.8b00683} {\bibfield  {journal} {\bibinfo
  {journal} {Nano Lett.}\ }\textbf {\bibinfo {volume} {18}},\ \bibinfo {pages}
  {3125} (\bibinfo {year} {2018})}\BibitemShut {NoStop}%
\bibitem [{\citenamefont {Chandra}\ and\ \citenamefont
  {Islam}(2010)}]{chandraElasticPropertiesMono2010}%
  \BibitemOpen
  \bibfield  {author} {\bibinfo {author} {\bibfnamefont {S.}~\bibnamefont
  {Chandra}}\ and\ \bibinfo {author} {\bibfnamefont {A.~K. M.~A.}\ \bibnamefont
  {Islam}},\ }\href {\doibase 10.1016/j.physc.2010.10.001} {\bibfield
  {journal} {\bibinfo  {journal} {Physica C Supercond}\ }\textbf {\bibinfo
  {volume} {470}},\ \bibinfo {pages} {2072} (\bibinfo {year}
  {2010})}\BibitemShut {NoStop}%
\bibitem [{\citenamefont {G{\'o}mez}\ \emph {et~al.}(2010)\citenamefont
  {G{\'o}mez}, \citenamefont {Marquina}, \citenamefont {{P{\'e}rez-Mazariego}},
  \citenamefont {Escamilla}, \citenamefont {Escudero}, \citenamefont
  {Quintana}, \citenamefont {{Hern{\'a}ndez-G{\'o}mez}}, \citenamefont
  {Ridaura},\ and\ \citenamefont {Marquina}}]{gomezEffectsSubstitutingSe2010}%
  \BibitemOpen
  \bibfield  {author} {\bibinfo {author} {\bibfnamefont {R.~W.}\ \bibnamefont
  {G{\'o}mez}}, \bibinfo {author} {\bibfnamefont {V.}~\bibnamefont {Marquina}},
  \bibinfo {author} {\bibfnamefont {J.~L.}\ \bibnamefont
  {{P{\'e}rez-Mazariego}}}, \bibinfo {author} {\bibfnamefont {R.}~\bibnamefont
  {Escamilla}}, \bibinfo {author} {\bibfnamefont {R.}~\bibnamefont {Escudero}},
  \bibinfo {author} {\bibfnamefont {M.}~\bibnamefont {Quintana}}, \bibinfo
  {author} {\bibfnamefont {J.~J.}\ \bibnamefont {{Hern{\'a}ndez-G{\'o}mez}}},
  \bibinfo {author} {\bibfnamefont {R.}~\bibnamefont {Ridaura}}, \ and\
  \bibinfo {author} {\bibfnamefont {M.~L.}\ \bibnamefont {Marquina}},\ }\href
  {\doibase 10.1007/s10948-010-0764-2} {\bibfield  {journal} {\bibinfo
  {journal} {J Supercond Nov Magn}\ }\textbf {\bibinfo {volume} {23}},\
  \bibinfo {pages} {551} (\bibinfo {year} {2010})}\BibitemShut {NoStop}%
\bibitem [{\citenamefont {Ciechan}\ \emph {et~al.}(2013)\citenamefont
  {Ciechan}, \citenamefont {Winiarski},\ and\ \citenamefont
  {{Samsel-Czeka{\l}a}}}]{ciechanMagneticPhaseTransitions2013}%
  \BibitemOpen
  \bibfield  {author} {\bibinfo {author} {\bibfnamefont {A.}~\bibnamefont
  {Ciechan}}, \bibinfo {author} {\bibfnamefont {M.~J.}\ \bibnamefont
  {Winiarski}}, \ and\ \bibinfo {author} {\bibfnamefont {M.}~\bibnamefont
  {{Samsel-Czeka{\l}a}}},\ }\href {\doibase 10.1088/0953-8984/26/2/025702}
  {\bibfield  {journal} {\bibinfo  {journal} {J. Phys.: Condens. Matter}\
  }\textbf {\bibinfo {volume} {26}},\ \bibinfo {pages} {025702} (\bibinfo
  {year} {2013})}\BibitemShut {NoStop}%
\bibitem [{\citenamefont {McGuire}\ \emph {et~al.}(2017)\citenamefont
  {McGuire}, \citenamefont {Garlea}, \citenamefont {KC}, \citenamefont
  {Cooper}, \citenamefont {Yan}, \citenamefont {Cao},\ and\ \citenamefont
  {Sales}}]{mcguireAntiferromagnetismVanWaals2017}%
  \BibitemOpen
  \bibfield  {author} {\bibinfo {author} {\bibfnamefont {M.~A.}\ \bibnamefont
  {McGuire}}, \bibinfo {author} {\bibfnamefont {V.~O.}\ \bibnamefont {Garlea}},
  \bibinfo {author} {\bibfnamefont {S.}~\bibnamefont {KC}}, \bibinfo {author}
  {\bibfnamefont {V.~R.}\ \bibnamefont {Cooper}}, \bibinfo {author}
  {\bibfnamefont {J.}~\bibnamefont {Yan}}, \bibinfo {author} {\bibfnamefont
  {H.}~\bibnamefont {Cao}}, \ and\ \bibinfo {author} {\bibfnamefont {B.~C.}\
  \bibnamefont {Sales}},\ }\href {\doibase 10.1103/PhysRevB.95.144421}
  {\bibfield  {journal} {\bibinfo  {journal} {Phys. Rev. B}\ }\textbf {\bibinfo
  {volume} {95}},\ \bibinfo {pages} {144421} (\bibinfo {year}
  {2017})}\BibitemShut {NoStop}%
\bibitem [{\citenamefont {Naumis}\ \emph {et~al.}(2017)\citenamefont {Naumis},
  \citenamefont {{Barraza-Lopez}}, \citenamefont {{Oliva-Leyva}},\ and\
  \citenamefont {Terrones}}]{naumisElectronicOpticalProperties2017}%
  \BibitemOpen
  \bibfield  {author} {\bibinfo {author} {\bibfnamefont {G.~G.}\ \bibnamefont
  {Naumis}}, \bibinfo {author} {\bibfnamefont {S.}~\bibnamefont
  {{Barraza-Lopez}}}, \bibinfo {author} {\bibfnamefont {M.}~\bibnamefont
  {{Oliva-Leyva}}}, \ and\ \bibinfo {author} {\bibfnamefont {H.}~\bibnamefont
  {Terrones}},\ }\href {\doibase 10.1088/1361-6633/aa74ef} {\bibfield
  {journal} {\bibinfo  {journal} {Rep. Prog. Phys.}\ }\textbf {\bibinfo
  {volume} {80}},\ \bibinfo {pages} {096501} (\bibinfo {year}
  {2017})}\BibitemShut {NoStop}%
\bibitem [{\citenamefont {Cao}\ \emph {et~al.}(2018)\citenamefont {Cao},
  \citenamefont {Fatemi}, \citenamefont {Fang}, \citenamefont {Watanabe},
  \citenamefont {Taniguchi}, \citenamefont {Kaxiras},\ and\ \citenamefont
  {{Jarillo-Herrero}}}]{caoUnconventionalSuperconductivityMagicangle2018}%
  \BibitemOpen
  \bibfield  {author} {\bibinfo {author} {\bibfnamefont {Y.}~\bibnamefont
  {Cao}}, \bibinfo {author} {\bibfnamefont {V.}~\bibnamefont {Fatemi}},
  \bibinfo {author} {\bibfnamefont {S.}~\bibnamefont {Fang}}, \bibinfo {author}
  {\bibfnamefont {K.}~\bibnamefont {Watanabe}}, \bibinfo {author}
  {\bibfnamefont {T.}~\bibnamefont {Taniguchi}}, \bibinfo {author}
  {\bibfnamefont {E.}~\bibnamefont {Kaxiras}}, \ and\ \bibinfo {author}
  {\bibfnamefont {P.}~\bibnamefont {{Jarillo-Herrero}}},\ }\href {\doibase
  10.1038/nature26160} {\bibfield  {journal} {\bibinfo  {journal} {Nature}\
  }\textbf {\bibinfo {volume} {556}},\ \bibinfo {pages} {43} (\bibinfo {year}
  {2018})}\BibitemShut {NoStop}%
\bibitem [{\citenamefont {Andrei}\ and\ \citenamefont
  {MacDonald}(2020)}]{andreiGrapheneBilayersTwist2020}%
  \BibitemOpen
  \bibfield  {author} {\bibinfo {author} {\bibfnamefont {E.~Y.}\ \bibnamefont
  {Andrei}}\ and\ \bibinfo {author} {\bibfnamefont {A.~H.}\ \bibnamefont
  {MacDonald}},\ }\href {\doibase 10.1038/s41563-020-00840-0} {\bibfield
  {journal} {\bibinfo  {journal} {Nat. Mater.}\ }\textbf {\bibinfo {volume}
  {19}},\ \bibinfo {pages} {1265} (\bibinfo {year} {2020})}\BibitemShut
  {NoStop}%
\bibitem [{\citenamefont {Xu}\ \emph {et~al.}(2021)\citenamefont {Xu},
  \citenamefont {Guo},\ and\ \citenamefont {Xian}}]{xuMoireFlatBands2021}%
  \BibitemOpen
  \bibfield  {author} {\bibinfo {author} {\bibfnamefont {Q.}~\bibnamefont
  {Xu}}, \bibinfo {author} {\bibfnamefont {Y.}~\bibnamefont {Guo}}, \ and\
  \bibinfo {author} {\bibfnamefont {L.}~\bibnamefont {Xian}},\ }\href {\doibase
  10.1088/2053-1583/ac3a98} {\bibfield  {journal} {\bibinfo  {journal} {2D
  Mater.}\ }\textbf {\bibinfo {volume} {9}},\ \bibinfo {pages} {014005}
  (\bibinfo {year} {2021})}\BibitemShut {NoStop}%
\bibitem [{\citenamefont {Vizner~Stern}\ \emph {et~al.}(2021)\citenamefont
  {Vizner~Stern}, \citenamefont {Waschitz}, \citenamefont {Cao}, \citenamefont
  {Nevo}, \citenamefont {Watanabe}, \citenamefont {Taniguchi}, \citenamefont
  {Sela}, \citenamefont {Urbakh}, \citenamefont {Hod},\ and\ \citenamefont
  {Ben~Shalom}}]{viznersternInterfacialFerroelectricityVan2021}%
  \BibitemOpen
  \bibfield  {author} {\bibinfo {author} {\bibfnamefont {M.}~\bibnamefont
  {Vizner~Stern}}, \bibinfo {author} {\bibfnamefont {Y.}~\bibnamefont
  {Waschitz}}, \bibinfo {author} {\bibfnamefont {W.}~\bibnamefont {Cao}},
  \bibinfo {author} {\bibfnamefont {I.}~\bibnamefont {Nevo}}, \bibinfo {author}
  {\bibfnamefont {K.}~\bibnamefont {Watanabe}}, \bibinfo {author}
  {\bibfnamefont {T.}~\bibnamefont {Taniguchi}}, \bibinfo {author}
  {\bibfnamefont {E.}~\bibnamefont {Sela}}, \bibinfo {author} {\bibfnamefont
  {M.}~\bibnamefont {Urbakh}}, \bibinfo {author} {\bibfnamefont
  {O.}~\bibnamefont {Hod}}, \ and\ \bibinfo {author} {\bibfnamefont
  {M.}~\bibnamefont {Ben~Shalom}},\ }\href {\doibase 10.1126/science.abe8177}
  {\bibfield  {journal} {\bibinfo  {journal} {Science}\ }\textbf {\bibinfo
  {volume} {372}},\ \bibinfo {pages} {1462} (\bibinfo {year}
  {2021})}\BibitemShut {NoStop}%
\bibitem [{\citenamefont {Wu}\ \emph {et~al.}(2023)\citenamefont {Wu},
  \citenamefont {Kang}, \citenamefont {Wu},\ and\ \citenamefont
  {Lin}}]{wuLargeSlidingRegulation2023}%
  \BibitemOpen
  \bibfield  {author} {\bibinfo {author} {\bibfnamefont {Q.}~\bibnamefont
  {Wu}}, \bibinfo {author} {\bibfnamefont {L.}~\bibnamefont {Kang}}, \bibinfo
  {author} {\bibfnamefont {J.}~\bibnamefont {Wu}}, \ and\ \bibinfo {author}
  {\bibfnamefont {Z.}~\bibnamefont {Lin}},\ }\href {\doibase
  10.1038/s41524-023-01127-z} {\bibfield  {journal} {\bibinfo  {journal} {npj
  Comput Mater}\ }\textbf {\bibinfo {volume} {9}},\ \bibinfo {pages} {1}
  (\bibinfo {year} {2023})}\BibitemShut {NoStop}%
\bibitem [{\citenamefont {Wu}\ and\ \citenamefont
  {Li}(2021)}]{wuSlidingFerroelectricity2D2021}%
  \BibitemOpen
  \bibfield  {author} {\bibinfo {author} {\bibfnamefont {M.}~\bibnamefont
  {Wu}}\ and\ \bibinfo {author} {\bibfnamefont {J.}~\bibnamefont {Li}},\ }\href
  {\doibase 10.1073/pnas.2115703118} {\bibfield  {journal} {\bibinfo  {journal}
  {PNAS}\ }\textbf {\bibinfo {volume} {118}},\ \bibinfo {pages} {e2115703118}
  (\bibinfo {year} {2021})}\BibitemShut {NoStop}%
\bibitem [{\citenamefont {Burch}\ \emph
  {et~al.}(2018{\natexlab{a}})\citenamefont {Burch}, \citenamefont {Mandrus},\
  and\ \citenamefont {Park}}]{burchMagnetismTwodimensionalVan2018c}%
  \BibitemOpen
  \bibfield  {author} {\bibinfo {author} {\bibfnamefont {K.~S.}\ \bibnamefont
  {Burch}}, \bibinfo {author} {\bibfnamefont {D.}~\bibnamefont {Mandrus}}, \
  and\ \bibinfo {author} {\bibfnamefont {J.-G.}\ \bibnamefont {Park}},\ }\href
  {\doibase 10.1038/s41586-018-0631-z} {\bibfield  {journal} {\bibinfo
  {journal} {Nature}\ }\textbf {\bibinfo {volume} {563}},\ \bibinfo {pages}
  {47} (\bibinfo {year} {2018}{\natexlab{a}})}\BibitemShut {NoStop}%
\bibitem [{\citenamefont {Fei}\ \emph {et~al.}(2018)\citenamefont {Fei},
  \citenamefont {Huang}, \citenamefont {Malinowski}, \citenamefont {Wang},
  \citenamefont {Song}, \citenamefont {Sanchez}, \citenamefont {Yao},
  \citenamefont {Xiao}, \citenamefont {Zhu}, \citenamefont {May}, \citenamefont
  {Wu}, \citenamefont {Cobden}, \citenamefont {Chu},\ and\ \citenamefont
  {Xu}}]{feiTwodimensionalItinerantFerromagnetism2018}%
  \BibitemOpen
  \bibfield  {author} {\bibinfo {author} {\bibfnamefont {Z.}~\bibnamefont
  {Fei}}, \bibinfo {author} {\bibfnamefont {B.}~\bibnamefont {Huang}}, \bibinfo
  {author} {\bibfnamefont {P.}~\bibnamefont {Malinowski}}, \bibinfo {author}
  {\bibfnamefont {W.}~\bibnamefont {Wang}}, \bibinfo {author} {\bibfnamefont
  {T.}~\bibnamefont {Song}}, \bibinfo {author} {\bibfnamefont {J.}~\bibnamefont
  {Sanchez}}, \bibinfo {author} {\bibfnamefont {W.}~\bibnamefont {Yao}},
  \bibinfo {author} {\bibfnamefont {D.}~\bibnamefont {Xiao}}, \bibinfo {author}
  {\bibfnamefont {X.}~\bibnamefont {Zhu}}, \bibinfo {author} {\bibfnamefont
  {A.~F.}\ \bibnamefont {May}}, \bibinfo {author} {\bibfnamefont
  {W.}~\bibnamefont {Wu}}, \bibinfo {author} {\bibfnamefont {D.~H.}\
  \bibnamefont {Cobden}}, \bibinfo {author} {\bibfnamefont {J.-H.}\
  \bibnamefont {Chu}}, \ and\ \bibinfo {author} {\bibfnamefont
  {X.}~\bibnamefont {Xu}},\ }\href {\doibase 10.1038/s41563-018-0149-7}
  {\bibfield  {journal} {\bibinfo  {journal} {Nature Mater}\ }\textbf {\bibinfo
  {volume} {17}},\ \bibinfo {pages} {778} (\bibinfo {year} {2018})}\BibitemShut
  {NoStop}%
\bibitem [{liu()}]{liusupplemental}%
  \BibitemOpen
  \href@noop {} {\ }\bibinfo {note} {See Supplemental Material for more
  detailed information on (I) First-principles calculation methods, (II)
  Symmetry analysis, (III) Robustness of altermagnetism, (IV) The relationship
  between the type of altermagnetism and monolayer symmetry, (V) The
  tight-binding model, (VI) The expression of spin current and spin Hall angle
  in the twisted altermagnets, (VII) The robustness of interlayer magnetic
  order, (VIII) Material realization, which inclueds
  Refs.\cite{kresseEfficientIterativeSchemes1996,perdewGeneralizedGradientApproximation1996a,marzariMaximallyLocalizedGeneralized1997,mostofiWannier90ToolObtaining2008,souzaMaximallyLocalizedWannier2001,freimuthSpinorbitTorquesCo2014c,MnBiTe2019_Nature,zhangTopologicalAxionStates2019prl,li2019intrinsicSA,burchMagnetismTwodimensionalVan2018b,cococcioniLinearResponseApproach2005}}\BibitemShut
  {NoStop}%
\bibitem [{\citenamefont {Bistritzer}\ and\ \citenamefont
  {MacDonald}(2011)}]{bistritzerMoireBandsTwisted2011}%
  \BibitemOpen
  \bibfield  {author} {\bibinfo {author} {\bibfnamefont {R.}~\bibnamefont
  {Bistritzer}}\ and\ \bibinfo {author} {\bibfnamefont {A.~H.}\ \bibnamefont
  {MacDonald}},\ }\href {\doibase 10.1073/pnas.1108174108} {\bibfield
  {journal} {\bibinfo  {journal} {PNAS}\ }\textbf {\bibinfo {volume} {108}},\
  \bibinfo {pages} {12233} (\bibinfo {year} {2011})}\BibitemShut {NoStop}%
\bibitem [{\citenamefont {Jung}\ \emph {et~al.}(2014)\citenamefont {Jung},
  \citenamefont {Raoux}, \citenamefont {Qiao},\ and\ \citenamefont
  {MacDonald}}]{jungInitioTheoryMoir2014}%
  \BibitemOpen
  \bibfield  {author} {\bibinfo {author} {\bibfnamefont {J.}~\bibnamefont
  {Jung}}, \bibinfo {author} {\bibfnamefont {A.}~\bibnamefont {Raoux}},
  \bibinfo {author} {\bibfnamefont {Z.}~\bibnamefont {Qiao}}, \ and\ \bibinfo
  {author} {\bibfnamefont {A.~H.}\ \bibnamefont {MacDonald}},\ }\href {\doibase
  10.1103/PhysRevB.89.205414} {\bibfield  {journal} {\bibinfo  {journal} {Phys.
  Rev. B}\ }\textbf {\bibinfo {volume} {89}},\ \bibinfo {pages} {205414}
  (\bibinfo {year} {2014})}\BibitemShut {NoStop}%
\bibitem [{\citenamefont {Miao}\ \emph {et~al.}(2018)\citenamefont {Miao},
  \citenamefont {Xu}, \citenamefont {Zhu}, \citenamefont {Zhou},\ and\
  \citenamefont {Sun}}]{miao2DIntrinsicFerromagnets2018b}%
  \BibitemOpen
  \bibfield  {author} {\bibinfo {author} {\bibfnamefont {N.}~\bibnamefont
  {Miao}}, \bibinfo {author} {\bibfnamefont {B.}~\bibnamefont {Xu}}, \bibinfo
  {author} {\bibfnamefont {L.}~\bibnamefont {Zhu}}, \bibinfo {author}
  {\bibfnamefont {J.}~\bibnamefont {Zhou}}, \ and\ \bibinfo {author}
  {\bibfnamefont {Z.}~\bibnamefont {Sun}},\ }\href {\doibase
  10.1021/jacs.7b12976} {\bibfield  {journal} {\bibinfo  {journal} {J. Am.
  Chem. Soc.}\ }\textbf {\bibinfo {volume} {140}},\ \bibinfo {pages} {2417}
  (\bibinfo {year} {2018})}\BibitemShut {NoStop}%
\bibitem [{\citenamefont {Co\"{\i}c}\ \emph {et~al.}(1981)\citenamefont
  {Co\"{\i}c}, \citenamefont {Spiesser}, \citenamefont {Palvadeau},\ and\
  \citenamefont {Rouxel}}]{coicChromiumIIIOxyhalides1981}%
  \BibitemOpen
  \bibfield  {author} {\bibinfo {author} {\bibfnamefont {L.}~\bibnamefont
  {Co\"{\i}c}}, \bibinfo {author} {\bibfnamefont {M.}~\bibnamefont {Spiesser}},
  \bibinfo {author} {\bibfnamefont {P.}~\bibnamefont {Palvadeau}}, \ and\
  \bibinfo {author} {\bibfnamefont {J.}~\bibnamefont {Rouxel}},\ }\href
  {\doibase 10.1016/0025-5408(81)90086-6} {\bibfield  {journal} {\bibinfo
  {journal} {MATER RES BULL}\ }\textbf {\bibinfo {volume} {16}},\ \bibinfo
  {pages} {229} (\bibinfo {year} {1981})}\BibitemShut {NoStop}%
\bibitem [{\citenamefont {Feng}\ \emph {et~al.}(2021)\citenamefont {Feng},
  \citenamefont {Peng}, \citenamefont {Dai}, \citenamefont {Huang},
  \citenamefont {Duan},\ and\ \citenamefont
  {Ma}}]{fengAntiferromagneticFerroelasticMultiferroics2021}%
  \BibitemOpen
  \bibfield  {author} {\bibinfo {author} {\bibfnamefont {Y.}~\bibnamefont
  {Feng}}, \bibinfo {author} {\bibfnamefont {R.}~\bibnamefont {Peng}}, \bibinfo
  {author} {\bibfnamefont {Y.}~\bibnamefont {Dai}}, \bibinfo {author}
  {\bibfnamefont {B.}~\bibnamefont {Huang}}, \bibinfo {author} {\bibfnamefont
  {L.}~\bibnamefont {Duan}}, \ and\ \bibinfo {author} {\bibfnamefont
  {Y.}~\bibnamefont {Ma}},\ }\href {\doibase 10.1063/5.0071685} {\bibfield
  {journal} {\bibinfo  {journal} {Appl. Phys. Lett}\ }\textbf {\bibinfo
  {volume} {119}},\ \bibinfo {pages} {173103} (\bibinfo {year}
  {2021})}\BibitemShut {NoStop}%
\bibitem [{\citenamefont {Wiedenmann}\ \emph {et~al.}(1983)\citenamefont
  {Wiedenmann}, \citenamefont {Venien}, \citenamefont {Palvadeau},\ and\
  \citenamefont
  {{Rossat-Mignod}}}]{wiedenmannMagneticOrderingQuasitwodimensional1983}%
  \BibitemOpen
  \bibfield  {author} {\bibinfo {author} {\bibfnamefont {A.}~\bibnamefont
  {Wiedenmann}}, \bibinfo {author} {\bibfnamefont {J.~P.}\ \bibnamefont
  {Venien}}, \bibinfo {author} {\bibfnamefont {P.}~\bibnamefont {Palvadeau}}, \
  and\ \bibinfo {author} {\bibfnamefont {J.}~\bibnamefont {{Rossat-Mignod}}},\
  }\href {\doibase 10.1088/0022-3719/16/27/016} {\bibfield  {journal} {\bibinfo
   {journal} {J. Phys. C: Solid State Phys.}\ }\textbf {\bibinfo {volume}
  {16}},\ \bibinfo {pages} {5339} (\bibinfo {year} {1983})}\BibitemShut
  {NoStop}%
\bibitem [{\citenamefont {Shen}\ \emph {et~al.}(2022)\citenamefont {Shen},
  \citenamefont {Su},\ and\ \citenamefont
  {He}}]{shenHighthroughputComputationStructure2022}%
  \BibitemOpen
  \bibfield  {author} {\bibinfo {author} {\bibfnamefont {Z.-X.}\ \bibnamefont
  {Shen}}, \bibinfo {author} {\bibfnamefont {C.}~\bibnamefont {Su}}, \ and\
  \bibinfo {author} {\bibfnamefont {L.}~\bibnamefont {He}},\ }\href {\doibase
  10.1038/s41524-022-00813-8} {\bibfield  {journal} {\bibinfo  {journal} {npj
  Comput Mater}\ }\textbf {\bibinfo {volume} {8}},\ \bibinfo {pages} {1}
  (\bibinfo {year} {2022})}\BibitemShut {NoStop}%
\bibitem [{\citenamefont {Zhang}\ \emph {et~al.}(2021)\citenamefont {Zhang},
  \citenamefont {Xu}, \citenamefont {Luo},\ and\ \citenamefont
  {Zou}}]{zhangTwodimensionalMagneticMaterials2021a}%
  \BibitemOpen
  \bibfield  {author} {\bibinfo {author} {\bibfnamefont {S.}~\bibnamefont
  {Zhang}}, \bibinfo {author} {\bibfnamefont {R.}~\bibnamefont {Xu}}, \bibinfo
  {author} {\bibfnamefont {N.}~\bibnamefont {Luo}}, \ and\ \bibinfo {author}
  {\bibfnamefont {X.}~\bibnamefont {Zou}},\ }\href {\doibase
  10.1039/D0NR06813F} {\bibfield  {journal} {\bibinfo  {journal} {Nanoscale}\
  }\textbf {\bibinfo {volume} {13}},\ \bibinfo {pages} {1398} (\bibinfo {year}
  {2021})}\BibitemShut {NoStop}%
\bibitem [{\citenamefont {Torelli}\ \emph {et~al.}(2020)\citenamefont
  {Torelli}, \citenamefont {Moustafa}, \citenamefont {Jacobsen},\ and\
  \citenamefont {Olsen}}]{torelliHighthroughputComputationalScreening2020}%
  \BibitemOpen
  \bibfield  {author} {\bibinfo {author} {\bibfnamefont {D.}~\bibnamefont
  {Torelli}}, \bibinfo {author} {\bibfnamefont {H.}~\bibnamefont {Moustafa}},
  \bibinfo {author} {\bibfnamefont {K.~W.}\ \bibnamefont {Jacobsen}}, \ and\
  \bibinfo {author} {\bibfnamefont {T.}~\bibnamefont {Olsen}},\ }\href
  {\doibase 10.1038/s41524-020-00428-x} {\bibfield  {journal} {\bibinfo
  {journal} {npj Comput Mater}\ }\textbf {\bibinfo {volume} {6}},\ \bibinfo
  {pages} {1} (\bibinfo {year} {2020})}\BibitemShut {NoStop}%
\bibitem [{\citenamefont {Zhang}\ \emph {et~al.}(2017)\citenamefont {Zhang},
  \citenamefont {Pang}, \citenamefont {Zhang}, \citenamefont {Gu},\ and\
  \citenamefont {Huang}}]{zhangTwoDimensionalCo2S2Monolayer2017}%
  \BibitemOpen
  \bibfield  {author} {\bibinfo {author} {\bibfnamefont {Y.}~\bibnamefont
  {Zhang}}, \bibinfo {author} {\bibfnamefont {J.}~\bibnamefont {Pang}},
  \bibinfo {author} {\bibfnamefont {M.}~\bibnamefont {Zhang}}, \bibinfo
  {author} {\bibfnamefont {X.}~\bibnamefont {Gu}}, \ and\ \bibinfo {author}
  {\bibfnamefont {L.}~\bibnamefont {Huang}},\ }\href {\doibase
  10.1038/s41598-017-16032-x} {\bibfield  {journal} {\bibinfo  {journal} {Sci
  Rep}\ }\textbf {\bibinfo {volume} {7}},\ \bibinfo {pages} {15993} (\bibinfo
  {year} {2017})}\BibitemShut {NoStop}%
\bibitem [{\citenamefont {Freimuth}\ \emph {et~al.}(2014)\citenamefont
  {Freimuth}, \citenamefont {Bl{\"u}gel},\ and\ \citenamefont
  {Mokrousov}}]{freimuthSpinorbitTorquesCo2014c}%
  \BibitemOpen
  \bibfield  {author} {\bibinfo {author} {\bibfnamefont {F.}~\bibnamefont
  {Freimuth}}, \bibinfo {author} {\bibfnamefont {S.}~\bibnamefont
  {Bl{\"u}gel}}, \ and\ \bibinfo {author} {\bibfnamefont {Y.}~\bibnamefont
  {Mokrousov}},\ }\href {\doibase 10.1103/PhysRevB.90.174423} {\bibfield
  {journal} {\bibinfo  {journal} {Phys. Rev. B}\ }\textbf {\bibinfo {volume}
  {90}},\ \bibinfo {pages} {174423} (\bibinfo {year} {2014})}\BibitemShut
  {NoStop}%
\bibitem [{\citenamefont {Seemann}\ \emph {et~al.}(2015)\citenamefont
  {Seemann}, \citenamefont {K{\"o}dderitzsch}, \citenamefont {Wimmer},\ and\
  \citenamefont {Ebert}}]{seemannSymmetryimposedShapeLinear2015}%
  \BibitemOpen
  \bibfield  {author} {\bibinfo {author} {\bibfnamefont {M.}~\bibnamefont
  {Seemann}}, \bibinfo {author} {\bibfnamefont {D.}~\bibnamefont
  {K{\"o}dderitzsch}}, \bibinfo {author} {\bibfnamefont {S.}~\bibnamefont
  {Wimmer}}, \ and\ \bibinfo {author} {\bibfnamefont {H.}~\bibnamefont
  {Ebert}},\ }\href {\doibase 10.1103/PhysRevB.92.155138} {\bibfield  {journal}
  {\bibinfo  {journal} {Phys. Rev. B}\ }\textbf {\bibinfo {volume} {92}},\
  \bibinfo {pages} {155138} (\bibinfo {year} {2015})}\BibitemShut {NoStop}%
\bibitem [{\citenamefont {Mosendz}\ \emph {et~al.}(2010)\citenamefont
  {Mosendz}, \citenamefont {Pearson}, \citenamefont {Fradin}, \citenamefont
  {Bauer}, \citenamefont {Bader},\ and\ \citenamefont
  {Hoffmann}}]{mosendzQuantifyingSpinHall2010}%
  \BibitemOpen
  \bibfield  {author} {\bibinfo {author} {\bibfnamefont {O.}~\bibnamefont
  {Mosendz}}, \bibinfo {author} {\bibfnamefont {J.~E.}\ \bibnamefont
  {Pearson}}, \bibinfo {author} {\bibfnamefont {F.~Y.}\ \bibnamefont {Fradin}},
  \bibinfo {author} {\bibfnamefont {G.~E.~W.}\ \bibnamefont {Bauer}}, \bibinfo
  {author} {\bibfnamefont {S.~D.}\ \bibnamefont {Bader}}, \ and\ \bibinfo
  {author} {\bibfnamefont {A.}~\bibnamefont {Hoffmann}},\ }\href {\doibase
  10.1103/PhysRevLett.104.046601} {\bibfield  {journal} {\bibinfo  {journal}
  {Phys. Rev. Lett.}\ }\textbf {\bibinfo {volume} {104}},\ \bibinfo {pages}
  {046601} (\bibinfo {year} {2010})}\BibitemShut {NoStop}%
\bibitem [{\citenamefont {Isasa}\ \emph {et~al.}(2015)\citenamefont {Isasa},
  \citenamefont {Villamor}, \citenamefont {Hueso}, \citenamefont {Gradhand},\
  and\ \citenamefont {Casanova}}]{isasaTemperatureDependenceSpin2015a}%
  \BibitemOpen
  \bibfield  {author} {\bibinfo {author} {\bibfnamefont {M.}~\bibnamefont
  {Isasa}}, \bibinfo {author} {\bibfnamefont {E.}~\bibnamefont {Villamor}},
  \bibinfo {author} {\bibfnamefont {L.~E.}\ \bibnamefont {Hueso}}, \bibinfo
  {author} {\bibfnamefont {M.}~\bibnamefont {Gradhand}}, \ and\ \bibinfo
  {author} {\bibfnamefont {F.}~\bibnamefont {Casanova}},\ }\href {\doibase
  10.1103/PhysRevB.91.024402} {\bibfield  {journal} {\bibinfo  {journal} {Phys.
  Rev. B}\ }\textbf {\bibinfo {volume} {91}},\ \bibinfo {pages} {024402}
  (\bibinfo {year} {2015})}\BibitemShut {NoStop}%
\bibitem [{\citenamefont {Karube}\ \emph {et~al.}(2022)\citenamefont {Karube},
  \citenamefont {Tanaka}, \citenamefont {Sugawara}, \citenamefont {Kadoguchi},
  \citenamefont {Kohda},\ and\ \citenamefont
  {Nitta}}]{karubeObservationSpinSplitterTorque2022}%
  \BibitemOpen
  \bibfield  {author} {\bibinfo {author} {\bibfnamefont {S.}~\bibnamefont
  {Karube}}, \bibinfo {author} {\bibfnamefont {T.}~\bibnamefont {Tanaka}},
  \bibinfo {author} {\bibfnamefont {D.}~\bibnamefont {Sugawara}}, \bibinfo
  {author} {\bibfnamefont {N.}~\bibnamefont {Kadoguchi}}, \bibinfo {author}
  {\bibfnamefont {M.}~\bibnamefont {Kohda}}, \ and\ \bibinfo {author}
  {\bibfnamefont {J.}~\bibnamefont {Nitta}},\ }\href {\doibase
  10.1103/PhysRevLett.129.137201} {\bibfield  {journal} {\bibinfo  {journal}
  {Phys. Rev. Lett.}\ }\textbf {\bibinfo {volume} {129}},\ \bibinfo {pages}
  {137201} (\bibinfo {year} {2022})}\BibitemShut {NoStop}%
\bibitem [{\citenamefont {Obstbaum}\ \emph {et~al.}(2016)\citenamefont
  {Obstbaum}, \citenamefont {Decker}, \citenamefont {Greitner}, \citenamefont
  {Haertinger}, \citenamefont {Meier}, \citenamefont {Kronseder}, \citenamefont
  {Chadova}, \citenamefont {Wimmer}, \citenamefont {K{\"o}dderitzsch},
  \citenamefont {Ebert},\ and\ \citenamefont
  {Back}}]{obstbaumTuningSpinHall2016}%
  \BibitemOpen
  \bibfield  {author} {\bibinfo {author} {\bibfnamefont {M.}~\bibnamefont
  {Obstbaum}}, \bibinfo {author} {\bibfnamefont {M.}~\bibnamefont {Decker}},
  \bibinfo {author} {\bibfnamefont {A.~K.}\ \bibnamefont {Greitner}}, \bibinfo
  {author} {\bibfnamefont {M.}~\bibnamefont {Haertinger}}, \bibinfo {author}
  {\bibfnamefont {T.~N.~G.}\ \bibnamefont {Meier}}, \bibinfo {author}
  {\bibfnamefont {M.}~\bibnamefont {Kronseder}}, \bibinfo {author}
  {\bibfnamefont {K.}~\bibnamefont {Chadova}}, \bibinfo {author} {\bibfnamefont
  {S.}~\bibnamefont {Wimmer}}, \bibinfo {author} {\bibfnamefont
  {D.}~\bibnamefont {K{\"o}dderitzsch}}, \bibinfo {author} {\bibfnamefont
  {H.}~\bibnamefont {Ebert}}, \ and\ \bibinfo {author} {\bibfnamefont {C.~H.}\
  \bibnamefont {Back}},\ }\href {\doibase 10.1103/PhysRevLett.117.167204}
  {\bibfield  {journal} {\bibinfo  {journal} {Phys. Rev. Lett.}\ }\textbf
  {\bibinfo {volume} {117}},\ \bibinfo {pages} {167204} (\bibinfo {year}
  {2016})}\BibitemShut {NoStop}%
\bibitem [{\citenamefont {Singh}\ and\ \citenamefont
  {Bedanta}(2020)}]{singhLargeSpinHall2020}%
  \BibitemOpen
  \bibfield  {author} {\bibinfo {author} {\bibfnamefont {B.~B.}\ \bibnamefont
  {Singh}}\ and\ \bibinfo {author} {\bibfnamefont {S.}~\bibnamefont
  {Bedanta}},\ }\href {\doibase 10.1103/PhysRevApplied.13.044020} {\bibfield
  {journal} {\bibinfo  {journal} {Phys. Rev. Appl.}\ }\textbf {\bibinfo
  {volume} {13}},\ \bibinfo {pages} {044020} (\bibinfo {year}
  {2020})}\BibitemShut {NoStop}%
\bibitem [{\citenamefont {Wang}\ \emph {et~al.}(2014)\citenamefont {Wang},
  \citenamefont {Deorani}, \citenamefont {Qiu}, \citenamefont {Kwon},\ and\
  \citenamefont {Yang}}]{wangDeterminationIntrinsicSpin2014}%
  \BibitemOpen
  \bibfield  {author} {\bibinfo {author} {\bibfnamefont {Y.}~\bibnamefont
  {Wang}}, \bibinfo {author} {\bibfnamefont {P.}~\bibnamefont {Deorani}},
  \bibinfo {author} {\bibfnamefont {X.}~\bibnamefont {Qiu}}, \bibinfo {author}
  {\bibfnamefont {J.~H.}\ \bibnamefont {Kwon}}, \ and\ \bibinfo {author}
  {\bibfnamefont {H.}~\bibnamefont {Yang}},\ }\href {\doibase
  10.1063/1.4898593} {\bibfield  {journal} {\bibinfo  {journal} {Appl. Phys.
  Lett}\ }\textbf {\bibinfo {volume} {105}},\ \bibinfo {pages} {152412}
  (\bibinfo {year} {2014})}\BibitemShut {NoStop}%
\bibitem [{\citenamefont {Kresse}\ and\ \citenamefont
  {Furthm{\"u}ller}(1996)}]{kresseEfficientIterativeSchemes1996}%
  \BibitemOpen
  \bibfield  {author} {\bibinfo {author} {\bibfnamefont {G.}~\bibnamefont
  {Kresse}}\ and\ \bibinfo {author} {\bibfnamefont {J.}~\bibnamefont
  {Furthm{\"u}ller}},\ }\href {\doibase 10.1103/PhysRevB.54.11169} {\bibfield
  {journal} {\bibinfo  {journal} {Phys. Rev. B}\ }\textbf {\bibinfo {volume}
  {54}},\ \bibinfo {pages} {11169} (\bibinfo {year} {1996})}\BibitemShut
  {NoStop}%
\bibitem [{\citenamefont {Perdew}\ \emph {et~al.}(1996)\citenamefont {Perdew},
  \citenamefont {Burke},\ and\ \citenamefont
  {Ernzerhof}}]{perdewGeneralizedGradientApproximation1996a}%
  \BibitemOpen
  \bibfield  {author} {\bibinfo {author} {\bibfnamefont {J.~P.}\ \bibnamefont
  {Perdew}}, \bibinfo {author} {\bibfnamefont {K.}~\bibnamefont {Burke}}, \
  and\ \bibinfo {author} {\bibfnamefont {M.}~\bibnamefont {Ernzerhof}},\ }\href
  {\doibase 10.1103/PhysRevLett.77.3865} {\bibfield  {journal} {\bibinfo
  {journal} {Phys. Rev. Lett.}\ }\textbf {\bibinfo {volume} {77}},\ \bibinfo
  {pages} {3865} (\bibinfo {year} {1996})}\BibitemShut {NoStop}%
\bibitem [{\citenamefont {Marzari}\ and\ \citenamefont
  {Vanderbilt}(1997)}]{marzariMaximallyLocalizedGeneralized1997}%
  \BibitemOpen
  \bibfield  {author} {\bibinfo {author} {\bibfnamefont {N.}~\bibnamefont
  {Marzari}}\ and\ \bibinfo {author} {\bibfnamefont {D.}~\bibnamefont
  {Vanderbilt}},\ }\href {\doibase 10.1103/PhysRevB.56.12847} {\bibfield
  {journal} {\bibinfo  {journal} {Phys. Rev. B}\ }\textbf {\bibinfo {volume}
  {56}},\ \bibinfo {pages} {12847} (\bibinfo {year} {1997})}\BibitemShut
  {NoStop}%
\bibitem [{\citenamefont {Mostofi}\ \emph {et~al.}(2008)\citenamefont
  {Mostofi}, \citenamefont {Yates}, \citenamefont {Lee}, \citenamefont {Souza},
  \citenamefont {Vanderbilt},\ and\ \citenamefont
  {Marzari}}]{mostofiWannier90ToolObtaining2008}%
  \BibitemOpen
  \bibfield  {author} {\bibinfo {author} {\bibfnamefont {A.~A.}\ \bibnamefont
  {Mostofi}}, \bibinfo {author} {\bibfnamefont {J.~R.}\ \bibnamefont {Yates}},
  \bibinfo {author} {\bibfnamefont {Y.-S.}\ \bibnamefont {Lee}}, \bibinfo
  {author} {\bibfnamefont {I.}~\bibnamefont {Souza}}, \bibinfo {author}
  {\bibfnamefont {D.}~\bibnamefont {Vanderbilt}}, \ and\ \bibinfo {author}
  {\bibfnamefont {N.}~\bibnamefont {Marzari}},\ }\href {\doibase
  10.1016/j.cpc.2007.11.016} {\bibfield  {journal} {\bibinfo  {journal} {Comput
  Phys Commun}\ }\textbf {\bibinfo {volume} {178}},\ \bibinfo {pages} {685}
  (\bibinfo {year} {2008})},\ \Eprint {http://arxiv.org/abs/0708.0650}
  {0708.0650} \BibitemShut {NoStop}%
\bibitem [{\citenamefont {Souza}\ \emph {et~al.}(2001)\citenamefont {Souza},
  \citenamefont {Marzari},\ and\ \citenamefont
  {Vanderbilt}}]{souzaMaximallyLocalizedWannier2001}%
  \BibitemOpen
  \bibfield  {author} {\bibinfo {author} {\bibfnamefont {I.}~\bibnamefont
  {Souza}}, \bibinfo {author} {\bibfnamefont {N.}~\bibnamefont {Marzari}}, \
  and\ \bibinfo {author} {\bibfnamefont {D.}~\bibnamefont {Vanderbilt}},\
  }\href {\doibase 10.1103/PhysRevB.65.035109} {\bibfield  {journal} {\bibinfo
  {journal} {Phys. Rev. B}\ }\textbf {\bibinfo {volume} {65}},\ \bibinfo
  {pages} {035109} (\bibinfo {year} {2001})}\BibitemShut {NoStop}%
\bibitem [{\citenamefont {Burch}\ \emph
  {et~al.}(2018{\natexlab{b}})\citenamefont {Burch}, \citenamefont {Mandrus},\
  and\ \citenamefont {Park}}]{burchMagnetismTwodimensionalVan2018b}%
  \BibitemOpen
  \bibfield  {author} {\bibinfo {author} {\bibfnamefont {K.~S.}\ \bibnamefont
  {Burch}}, \bibinfo {author} {\bibfnamefont {D.}~\bibnamefont {Mandrus}}, \
  and\ \bibinfo {author} {\bibfnamefont {J.-G.}\ \bibnamefont {Park}},\ }\href
  {\doibase 10.1038/s41586-018-0631-z} {\bibfield  {journal} {\bibinfo
  {journal} {Nature}\ }\textbf {\bibinfo {volume} {563}},\ \bibinfo {pages}
  {47} (\bibinfo {year} {2018}{\natexlab{b}})}\BibitemShut {NoStop}%
\bibitem [{\citenamefont {Cococcioni}\ and\ \citenamefont {{de
  Gironcoli}}(2005)}]{cococcioniLinearResponseApproach2005}%
  \BibitemOpen
  \bibfield  {author} {\bibinfo {author} {\bibfnamefont {M.}~\bibnamefont
  {Cococcioni}}\ and\ \bibinfo {author} {\bibfnamefont {S.}~\bibnamefont {{de
  Gironcoli}}},\ }\href {\doibase 10.1103/PhysRevB.71.035105} {\bibfield
  {journal} {\bibinfo  {journal} {Phys. Rev. B}\ }\textbf {\bibinfo {volume}
  {71}},\ \bibinfo {pages} {035105} (\bibinfo {year} {2005})}\BibitemShut
  {NoStop}%
\end{thebibliography}%

\end{document}